\documentclass[12pt,preprint]{aastex}
\def\hal{H$\alpha$}

\def\be{\begin{equation}}
\def\ee{\end{equation}}

\def\m{~$\mu$m}

\def\HI   {\ion{H}{1}}
\def\HII  {\ion{H}{2}}

\def\ISO{{\it ISO}}
\def\IRAS{{\it IRAS}}

\begin {document}

\title{Herschel Far-Infrared and Sub-millimeter Photometry for the KINGFISH Sample of Nearby Galaxies}
\shorttitle{KINGFISH Photometry}

\author {
D.~A. Dale\altaffilmark{1},
G. Aniano\altaffilmark{2},
C.~W. Engelbracht\altaffilmark{3},
J.~L. Hinz\altaffilmark{3},
O. Krause\altaffilmark{4}, 
E.~J. Montiel\altaffilmark{3},
H. Roussel\altaffilmark{5},
P.~N. Appleton\altaffilmark{6},
L. Armus\altaffilmark{7},
P. Beir\~{a}o\altaffilmark{7},
A.~D. Bolatto\altaffilmark{8},
B.~R. Brandl\altaffilmark{9},
D. Calzetti\altaffilmark{10},
A.~F. Crocker\altaffilmark{10},
K.~V. Croxall\altaffilmark{11},
B.~T. Draine\altaffilmark{2},
M. Galametz\altaffilmark{12},
K.~D. Gordon\altaffilmark{13},
B.~A. Groves\altaffilmark{4},
C.-N. Hao\altaffilmark{14},
G. Helou\altaffilmark{6},
L.~K. Hunt\altaffilmark{15},
B.~D. Johnson\altaffilmark{5},
R.~C. Kennicutt\altaffilmark{12},
J. Koda\altaffilmark{16},
A.~K. Leroy\altaffilmark{17},
Y. Li\altaffilmark{10},
S.~E. Meidt\altaffilmark{4},
A.~E. Miller\altaffilmark{1},
E.~J. Murphy\altaffilmark{7},
N. Rahman\altaffilmark{8},
H.-W. Rix\altaffilmark{4},
K.~M. Sandstrom\altaffilmark{4}, 
M. Sauvage\altaffilmark{18},
E. Schinnerer\altaffilmark{4},
R.~A. Skibba\altaffilmark{3},
J.-D.~T. Smith\altaffilmark{11},
F.S. Tabatabaei\altaffilmark{4},
F. Walter\altaffilmark{4},
C.~D. Wilson\altaffilmark{19},
M.~G. Wolfire\altaffilmark{8}, and
S. Zibetti\altaffilmark{20}
}
\altaffiltext{1}{Department of Physics \& Astronomy, University of Wyoming, Laramie, WY 82071, USA; ddale@uwyo.edu}
\altaffiltext{2}{Department of Astrophysical Sciences, Princeton University, Princeton, NJ 08544, USA}
\altaffiltext{3}{Steward Observatory, University of Arizona, Tucson, AZ 85721, USA}
\altaffiltext{4}{Max-Planck-Institut f\"ur Astronomie, K\"onigstuhl 17, D-69117 Heidelberg, Germany}
\altaffiltext{5}{Institut d'Astrophysique de Paris, UMR7095 CNRS, Universit\'e Pierre \& Marie Curie, 98 bis Boulevard Arago, 75014 Paris, France}
\altaffiltext{6}{NASA Herschel Science Center, IPAC, California Institute of Technology, Pasadena, CA 91125, USA}
\altaffiltext{7}{Spitzer Science Center, California Institute of Technology, MC 314-6, Pasadena, CA 91125, USA}
\altaffiltext{8}{Department of Astronomy, University of Maryland, College Park, MD 20742, USA}
\altaffiltext{9}{Leiden Observatory, Leiden University, P.O. Box 9513, 2300 RA Leiden, The Netherlands}
\altaffiltext{10}{Department of Astronomy, University of Massachusetts, Amherst, MA 01003, USA}
\altaffiltext{11}{Department of Physics and Astronomy, University of Toledo, Toledo, OH 43606, USA}
\altaffiltext{12}{Institute of Astronomy, University of Cambridge, Madingley Road, Cambridge CB3 0HA, UK}
\altaffiltext{13}{Space Telescope Science Institute, 3700 San Martin Drive, Baltimore, MD 21218, USA}
\altaffiltext{14}{Tianjin Astrophysics Center, Tianjin Normal University, Tianjin 300387, China}
\altaffiltext{15}{INAF - Osservatorio Astrofisico di Arcetri, Largo E. Fermi 5, 50125 Firenze, Italy}
\altaffiltext{16}{Department of Physics and Astronomy, SUNY Stony Brook, Stony Brook, NY 11794-3800, USA}
\altaffiltext{17}{National Radio Astronomy Observatory, 520 Edgemont Road, Charlottesville, VA 22903, USA}
\altaffiltext{18}{CEA/DSM/DAPNIA/Service d'Astrophysique, UMR AIM, CE Saclay, 91191 Gif sur Yvette Cedex, France}
\altaffiltext{19}{Department of Physics \& Astronomy, McMaster University, Hamilton, Ontario L8S 4M1, Canada}
\altaffiltext{20}{Dark Cosmology Centre, Niels Bohr Institute, University of Copenhagen, Juliane Maries Vej 30, DK-2100 Copenhagen, Denmark}

\begin {abstract}
New far-infrared and sub-millimeter photometry from the Herschel Space Observatory is presented for 61 nearby galaxies from the Key Insights on Nearby Galaxies: A Far-Infrared Survey with Herschel (KINGFISH) sample.  The spatially-integrated fluxes are largely consistent with expectations based on Spitzer far-infrared photometry and extrapolations to longer wavelengths using popular dust emission models.  Dwarf irregular galaxies are notable exceptions, as already noted by other authors, as their 500\m\ emission shows evidence for a sub-millimeter excess.  In addition, the fraction of dust heating attributed to intense radiation fields associated with photo-dissociation regions is found to be ($21\pm4$)\% larger when Herschel data are included in the analysis. 
Dust masses obtained from the dust emission models of Draine \& Li are found to be on average nearly a factor of two higher than those based on single-temperature modified blackbodies, as single blackbody curves do not capture the full range of dust temperatures inherent to any galaxy.  The discrepancy is largest for galaxies exhibiting the coolest far-infrared colors.

\end {abstract}



\section {Introduction}
\label{sec:intro}



The amount of dust contained within a galaxy reflects the integrated chemical enrichment of the interstellar medium through accumulated episodes of star formation and any merger/accretion events, coupled with the galaxy's history of dust grain formation and destruction.
Thus, the total amount of dust observed within a galaxy can be tied to its star formation history.  From an observational viewpoint, estimating a galaxy's dust mass depends critically on far-infrared/sub-millimeter\footnote{``Far-infrared'' and ``sub-millimeter'' are defined here as $40 \leq \lambda \leq 300$ and $300 \leq \lambda \leq 1000~\micron$, respectively.} photometry and the inferred distribution of dust grain temperatures; long wavelength data are crucial for probing cool 15--20~K dust with any accuracy, and dust emitting at this temperature range makes up the bulk of the dust mass in a typical star-forming galaxy \citep{dunne01}.
Recent observational efforts have shown evidence for excess emission at sub-millimeter wavelengths in galaxies, particularly in low-metallicity systems such as the Small Magellanic Cloud and other dwarf galaxies \citep[e.g.,][]{bolatto00,galliano05,bot10,gordon10,ade11a,ade11b,galametz11,galliano11}.  If the excess emission is interpreted as evidence for particularly cold dust (e.g., $T \lesssim 10$~K), substantial amounts must be present in order to produce the observed emission, and the resulting dust-to-gas mass ratios inferred from these observations are significantly higher than what would be expected based on a galaxy's metallicity \citep[see, e.g.,][]{galametz09,meixner10,ade11a}.  Alternative explanations for the excess emission include resonances due to impurities in the dust or a modified dust emissivity at sub-millimeter wavelengths, e.g., changes with environment in dust grain size and/or composition \citep[see][and references therein]{lisenfeld02,aguirre03,dupac03,meny07,galametz09,meixner10}.

Understanding the physical origin of the sub-millimeter emission in nearby galaxies obviously requires sensitive sub-millimeter data.  However, it has been difficult to obtain reliable sub-millimeter photometry.  The opacity of the atmosphere limits ground-based observations to a few partially-transparent sub-millimeter spectral windows, and previous space-based missions such as {\it COBE}, {\it IRAS}, {\it ISO}, {\it Spitzer}, and {\it AKARI} lacked the sensitive instrumentation and the requisite angular resolution for resolved studies of nearby galaxies at these wavelengths.  The {\it Herschel Space Observatory} \citep{pilbratt10} allows an unprecedented look into the long wavelength emission from galaxies.  {\it Herschel} provides impressive angular resolution in the far-infrared/sub-millimeter wavelength range
and
unparalleled sensitivity to low surface brightness emission at 250, 350, and 500\m, wavelengths that bridge an important spectral gap between space-based galaxy surveys carried out at 160--170\m\ ({\it ISO}, {\it Spitzer}, {\it AKARI}) and those from the ground at 850--870\m; there are precious few detections of galaxies at 350 or 450\m\ of JCMT/SCUBA or CSO/SHARC heritage \cite[e.g.,][]{benford99,dunne01,kovacs06}.  Filling in this gap in wavelength coverage is particularly important for the coldest galaxies in the local universe, in which the dust emission peaks at $\sim$150\m\ or longer wavelengths.

In this effort we present {\it Herschel} PACS \citep{poglitsch10} and SPIRE \citep{griffin10} broadband photometry for the 61 galaxies in KINGFISH (Key Insights on Nearby Galaxies: A Far-Infrared Survey with {\it Herschel}).  The flux densities provided here are spatially-integrated or ``global'' values, and thus they do not directly take advantage of one of the key features of {\it Herschel} data: angular resolution.  However, the sharpness of the imaging comes into play by enabling superior identification and removal of contamination from neighboring and background galaxies.
While this study of global flux densities only indirectly profits from {\it Herschel}'s superior resolution, other studies investigate the spatially-resolved properties of KINGFISH galaxies.  For example, \cite{walter11}, \cite{aniano12}, 
and \cite{gordon12a} explore how the infrared spectral energy distributions vary with location in KINGFISH targets, \cite{galametz12}, \cite{hinz12}, and \cite{hunt12} study cold dust emission in the outer disks, \cite{li12} and \cite{crocker12} characterize the infrared properties of \HII\ regions, etc.

The global flux densities presented here are used to see how well the far-infrared/sub-millimeter emission in nearby galaxies can be modeled using the theoretical spectral energy distribution curves of \cite{draineli07}, curves that are based on the dust emission properties of the Milky Way.  We also quantify whether the sub-millimeter data provide new insight into galaxy spectral energy distributions or whether their values are expected based on extrapolations from shorter wavelength data.  Finally, we search for evidence for a global sub-millimeter excess in KINGFISH galaxies, explore the unique characteristics of the low-metallicity systems in the sample, and contrast the dust masses found through \cite{draineli07} fits with those from the common approach of fitting single-temperature modified blackbodies. 

Section~\ref{sec:sample} describes the sample and \S~\ref{sec:data} reviews the observations as well as the data processing and aperture photometry procedures that have been adopted here for KINGFISH imaging.  In \S~\ref{sec:results} the spatially-integrated photometry data are presented in addition to the results of fits to the observed spectral energy distributions, and \S~\ref{sec:summary} provides a summary of our findings.

\section {Sample}
\label{sec:sample}

The KINGFISH sample of galaxies draws from the {\it Spitzer} Infrared Nearby Galaxies Survey \citep[SINGS; see][]{kennicutt03}; fifty-seven SINGS galaxies are in the KINGFISH sample, plus NGC~2146, NGC~3077, NGC~5457, and IC~342.  The 61 galaxies were selected to span wide ranges in luminosity, optical/infrared ratio, and morphology \citep{kennicutt12}.  The luminosity ranges over a factor of $10^4$ (but all are sub-LIRG, or nearly so in the case of NGC~2146, with $L_{\rm IR} < 10^{11} L_\odot$), the range of optical/infrared ratios covers a factor of $10^3$, and all ``normal'' galaxy types are represented.  There are several galaxies with nuclei that are clearly distinguished by Seyfert or LINER characteristics, but none of the galaxies has their global luminosity dominated by an active nucleus \citep{moustakas10}.  The sample also spans representative ranges in metallicity, gas fraction, \HI/H$_2$ ratio, star formation rate, and bar strength.

\section {Observations and Data Processing}
\label{sec:data}

The observational program and data processing procedures for KINGFISH are described in detail in \cite{engelbracht10}, \cite{sandstrom10}, and \cite{kennicutt12}.  A brief description is provided here.  All 61 KINGFISH galaxies were imaged with PACS and SPIRE.

\subsection{PACS Data}

PACS imaging was obtained in scan mode, along two perpendicular axes for improved image reconstruction, at the medium scan speed of 20\arcsec\ s$^{-1}$.  The 45\degr\ orientation of the array with respect to the scan direction contributes to a more uniform spatial coverage.  Two Astronomical Observation Requests (AORs) were carried out for joint 70 and 160\m\ imaging, and an additional two AORs were utilized for joint 100 and 160\m\ observations, resulting in a total of four AORs for 160\m\ imaging.  Three or six repetitions were carried out for each AOR, depending on an individual galaxy's far-infrared surface brightness as gauged from {\it Spitzer}/MIPS data.  
The integrations achieved per pixel lead to 
approximate 1$\sigma$ surface brightness sensitivities of $\sigma_{\rm sky} \sim 5$, 5, and 2~MJy~sr$^{-1}$ at 70, 100, and 160\m, respectively, for the fainter subset of galaxies and approximately $\sqrt{2}$ times larger for the brighter subset.  The PACS calibration uncertainties are $\epsilon_{\rm cal,\nu}/f_\nu \sim 5$\%, according to Version~2.3 (08~June~2011) of the PACS Observer's Manual.



The raw (``Level~0'') data were processed using Version~5.0 of HIPE \citep{ott10}.  Besides the standard pipeline procedures, the conversion from Level~0 to Level~1 data included second-level deglitching and corrections for any offsets in the detector sub-matrices.  
Scanamorphos\footnote{http://www2.iap.fr/users/roussel/herschel} \citep{roussel11} Version 12.5 was used to process the Level~1 PACS scan map data.  Its main task is to subtract the brightness drifts caused by the low-frequency noise (comprising both the thermal drifts of the telescope and detectors and the flicker noise of the individual bolometers), before projecting the data onto a changeable spatial grid.  The algorithm employs minimal assumptions about the noise and the signal, and extracts the drifts from the data themselves, taking advantage of the redundancy built in to the scan observations.  With the nominal settings used by the KINGFISH survey, the drifts can be determined on timescales greater than or equal to 0.7~s at 70 and 100\m, and 0.9~s at 160\m\ (for a sampling interval of 0.1~s).  These timescales correspond to lengths between 1.5 and 2.5 times the beam FWHM, from 160\m\ to 70\m.  Second-level deglitching was performed, and the option to detect and mask brightness discontinuities was also used.  The data are weighted by the inverse square high-frequency noise of each bolometer in each scan.

The (``Level~2'') output of Scanamorphos is in the form of a FITS cube for each filter.  The four planes are the signal map, the error map, the map of the drifts that have been subtracted, and the weight map.  There is currently no propagation of the errors associated with the successive processing steps in the pipeline.  In each pixel the error is defined as the unbiased statistical estimate of the error on the mean.  The brightness unit is Jy~pixel$^{-1}$, and the pixel size is $\sim$ one-fourth of the beam FWHM, i.e., 1.4\arcsec\ at 70\m, 1.7\arcsec\ at 100\m, and 2.85\arcsec\ at 160\m.

\subsection{SPIRE Data}

SPIRE imaging data were taken in Large-Map mode, to an extent tailored to each galaxy's size (out to at least $\sim 1.5$ times the optical size).  Either two or four scans were obtained for each galaxy, based on its {\it Spitzer}/MIPS far-infrared surface brightness.  The resulting 1$\sigma$ limiting surface brightnesses are approximately $\sigma_{\rm sky} \sim 0.7$, 0.4, and 0.2~MJy~sr$^{-1}$ at 250, 350, and 500\m, respectively, for the fainter subset and $\sqrt{2}$ larger values for the brighter galaxies.
Calibration uncertainties for SPIRE data are estimated at $\epsilon_{\rm cal}/f_\nu \approx 7$\%, following Version~2.4 (07~June~2011) of the SPIRE Observer's Manual.  However, the uncertainties are strongly correlated between the three bands and thus the uncertainty on some SPIRE quantities such as the $f_\nu(250\mu m)/f_\nu(500\mu m)$ flux density ratio, for example, are less than the simplistic $\sqrt{2} \times 7$\% expectation.

SPIRE observations for six of our galaxies were obtained in the {\it Herschel} Reference Survey \citep{boselli10}: NGC~4254, NGC~4321, NGC~4536, NGC~4569, NGC~4579, and NGC~4725; those observations were not duplicated for KINGFISH.  The only notable difference between the SPIRE observations for the {\it Herschel} Reference Survey and those for KINGFISH is that three scans were employed (versus either two or four for KINGFISH observations, as described above).


The raw SPIRE data are processed through the early stages of HIPE (Version~5.0) to fit slopes to the data ramps and to calibrate the data in physical units.  A line is fit to the data for each scan leg after masking out the galaxy, and this fit is subtracted from the data.  Discrepant data (usually due to a rogue bolometer, of which there are $<1$ per map) are also masked, and the data are mosaicked using the mapper in HIPE.  The map coordinates are then adjusted so that the position of the point sources \citep[measured using StarFinder;][]{diolaiti00} match those in the MIPS 24~\micron\ images.  Finally, the images are converted to surface brightness units by dividing by the beam areas published in the SPIRE Observer's Manual:  423, 751, and 1587~$\sq$\arcsec\ at 250, 350, and 500\m, respectively.
Pixel sizes are 6\arcsec, 10\arcsec, and 14\arcsec\ at 250, 350, and 500\m, respectively.

\subsection{Sky Subtraction} 
\label{sec:skysub}

\begin{figure}
 \includegraphics[clip=true]{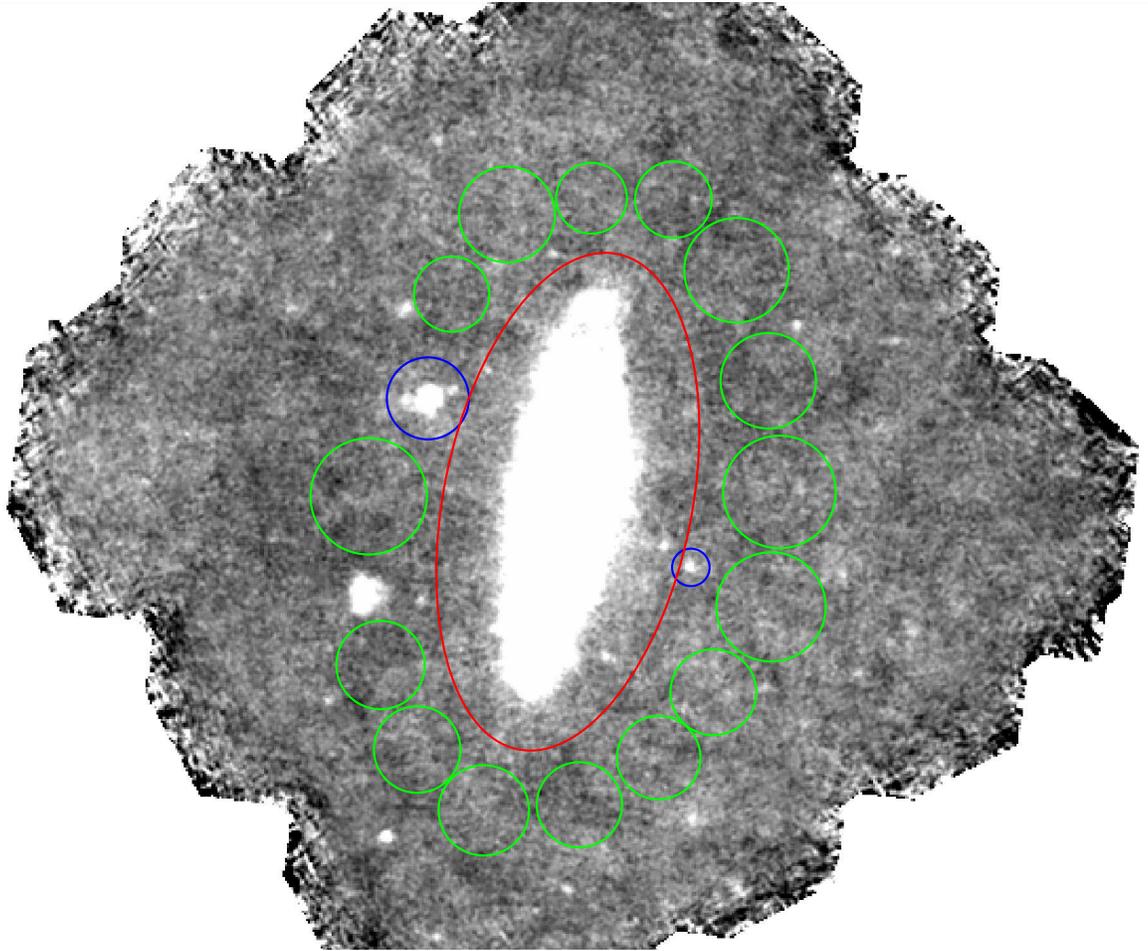}
 \caption{A PACS 160\m\ image of KINGFISH galaxy NGC~7331.  The large red ellipse indicates the photometric aperture listed in Table~\ref{tab:sample}, the green circles show the sky apertures, and the two blue circles identify sources to be removed before the photometry is executed.  North is up, East is to the left.}
 \label{fig:n7331}
\end{figure}

At far-infrared/sub-millimeter wavelengths the emission from the sky (above the atmosphere) largely comes from Milky Way cirrus and background galaxies.  However, the bolometer arrays of {\it Herschel} are not absolute photometers, and thus any map produced by any software is the superposition of an estimate of the true sky emission and an unknown (large) offset.  Hereafter this superposition is referred to as simply the ``sky''.  While the post-pipeline processed KINGFISH SPIRE and PACS images have their overall sky levels removed to zeroth order, a procedure has been adopted to remove a more refined local sky value for each galaxy.  To accomplish this local sky subtraction, for each PACS and SPIRE image a set of sky apertures has been defined that collectively circumscribe the galaxy, projected on the sky close enough to the galaxy to measure the ``local'' sky but far enough away to avoid containing any galaxy emission (Figure~\ref{fig:n7331}).  The emission from any prominent neighboring and/or background galaxies that are projected to lie within the sky apertures is removed before the sky is estimated.  
The total sky area, derived from the sum of the areas from all sky apertures, is typically significantly greater than that covered by the galaxy aperture itself, thereby limiting the contribution of uncertainty in the sky level to the overall error budget.  The mean sky level per pixel is computed from the collection of these sky apertures, the value is scaled to the number of pixels in the galaxy aperture, and the result is subtracted off from the overall galaxy aperture counts (all done within IRAF/{\tt IMCNTS}).  


\cite{aniano12} follow a different procedure for subtracting the sky emission from KINGFISH imaging, including fitting a tilted plane to the sky for each galaxy instead of a single value approach adopted here.  \cite{aniano12} study the spatial variations in the far-infrared/sub-millimeter spectral energy distributions and thus a more detailed characterization of the local sky is necessary.  The effects of most sky gradients cancel out in extracting
spatially integrated fluxes; the two approaches yield generally consistent global fluxes.

\subsection{Aperture Photometry}
\label{sec:ap_phot}

The elliptical apertures used for global photometry are listed in Table~\ref{tab:sample}.  The apertures are chosen by eye to encompass essentially all of the emission at every wavelength; aperture corrections described below are incorporated to recover the amount of any light that lies beyond these apertures.  The average ratio of aperture major axis length $2a$ to the de Vaucouleurs $D_{25}$ optical major axis is 1.45 (with a 1$\sigma$ dispersion in this ratio of 0.45).  The same aperture is used to extract the flux at each wavelength studied here, and they are very similar, if not identical, to those used for SINGS photometry \citep{dale07}, for the 57 galaxies that overlap the two samples.  As a test of the robustness of the aperture choices, the global flux densities using these apertures are compared to the values obtained using apertures with 5\% larger and 5\% smaller semi-minor and semi-major axes.  The impact of using $\sim$10\% larger or $\sim$10\% smaller aperture areas is a median difference of less than 1\% on the flux densities for all wavelengths.

Prior to extracting fluxes from aperture photometry, any emission from neighboring or background galaxies is identified and removed from the area covered by each aperture.  The identification is assisted by ancillary data at shorter wavelengths and higher spatial resolution (e.g., {\it Spitzer}/IRAC 3.6 and 8.0\m, {\it HST} optical, and ground-based \hal\ imaging).  The removal is accomplished via IRAF/{\tt IMEDIT} by replacing the values of contaminated pixels with the values from a random selection of nearby sky pixels, thereby incorporating the same noise statistics as the sky.  Usually the removal of such emission affects the global flux at less than the 1\% level, but in a few cases the impact is quite important, e.g., NGC~1317 lies within the aperture of NGC~1316, and background galaxies in the fields of the fainter dwarfs like Ho~II and DDO~053 would contribute significantly to the integrated flux (by up to $\sim 30-50\%$) if not removed \cite[see also][]{walter07}.

Diffraction inevitably results in a small portion of the galaxy emission appearing beyond the chosen apertures, however, and thus aperture corrections are formulated to mitigate this effect.  Aperture corrections are empirically determined from a comparison of fluxes from smoothed and unsmoothed {\it Spitzer}/IRAC 3.6\m\ imaging, which has a native resolution of $\sim$1\farcs5.  The aperture correction for a given PACS or SPIRE flux is the ratio of the flux from the unsmoothed 3.6\m\ image to the flux from the 3.6\m\ image smoothed to the same PSF as the {\it Herschel} image in question.  Due to the typically generous aperture size and sharp {\it Herschel} PACS and SPIRE PSFs, the amplitudes of the KINGFISH global photometry aperture corrections are typically quite small, with median values of 1.0 at all wavelengths and maximum values of 1.03 for PACS and between 1.07 and 1.13 for SPIRE.  This technique assumes that a galaxy's profile in the far-infrared matches that of its (mostly) stellar profile in the near-infrared, and there may in fact be appreciable differences in the two emission profiles.


The uncertainties in the integrated photometry $\epsilon_{\rm total}$ are computed as a combination in quadrature of the calibration uncertainty $\epsilon_{\rm cal}$ and the measurement uncertainty $\epsilon_{\rm sky}$ based on the measured sky fluctuations and the areas covered by the galaxy and the sum of the sky apertures, i.e., 
\be
\epsilon_{\rm total} = \sqrt{\epsilon_{\rm cal}^2 + \epsilon_{\rm sky}^2}
\ee
with
\be
\epsilon_{\rm sky} ~ = \sigma_{\rm sky} \Omega_{\rm pix} ~ \sqrt{N_{\rm pix}+{N_{\rm pix}}^2/N_{\rm sky}}
\ee
where $\sigma_{\rm sky}$ is the standard deviation of the sky surface brightness fluctuations,
$\Omega_{\rm pix}$ is the solid angle subtended per pixel, and $N_{\rm pix}$ and $N_{\rm sky}$ 
are the number of pixels in the galaxy and (the sum of) the sky apertures, respectively.  
For the few sources undetected by {\it Herschel} imaging, 5$\sigma$ upper limits are derived assuming a galaxy spans all $N_{\rm pix}$ pixels in the aperture,
\be
f_\nu(5\sigma~{\rm upper~limit}) ~ = ~ 5 ~ \epsilon_{\rm sky}.
\ee

\section {Results}
\label{sec:results}

\subsection {Flux Densities} 

Table~\ref{tab:data} presents the spatially-integrated flux densities for all 61 KINGFISH galaxies for all six {\it Herschel} photometric bands.  The tabulated flux densities include aperture corrections (\S~\ref{sec:ap_phot}) and have Galactic extinction \citep{schlegel98} removed assuming $A_V/E(B-V)\approx3.1$ and the reddening curve of \cite{lidraine01}.\footnote{The corrections for Galactic extinction are very small at these wavelengths, with the largest correction being 0.4\% for IC~0342 at 70\m, which lies at a Galactic latitude of $+10$\degr\ behind a foreground veil of $E(B-V)\sim0.56$~mag.}  No color corrections have been applied to the data in Table~\ref{tab:data}.  The most recent calibrations are used for both SPIRE and PACS photometry, including the ``FM,6'' PACS calibration which lowers the fluxes for extended sources by 10--20\% compared to the previous calibration, as described in the HIPE 7.0.0 documentation.

The superior sensitivity and angular resolution of {\it Herschel} enables a more detailed investigation of the faintest galaxies in our sample.  For example, \cite{dale07} provide marginally significant MIPS flux densities at 70 and 160\m\ for NGC~1404 and DDO~165, but they caution that the emission appearing within the apertures for these galaxies potentially derives from background galaxies.  It is now clear based on the {\it Herschel} maps that these targets were indeed not detected by {\it Spitzer} (nor {\it Herschel}) at $\lambda \geq 70$\m.

\begin{figure}
 \plotone{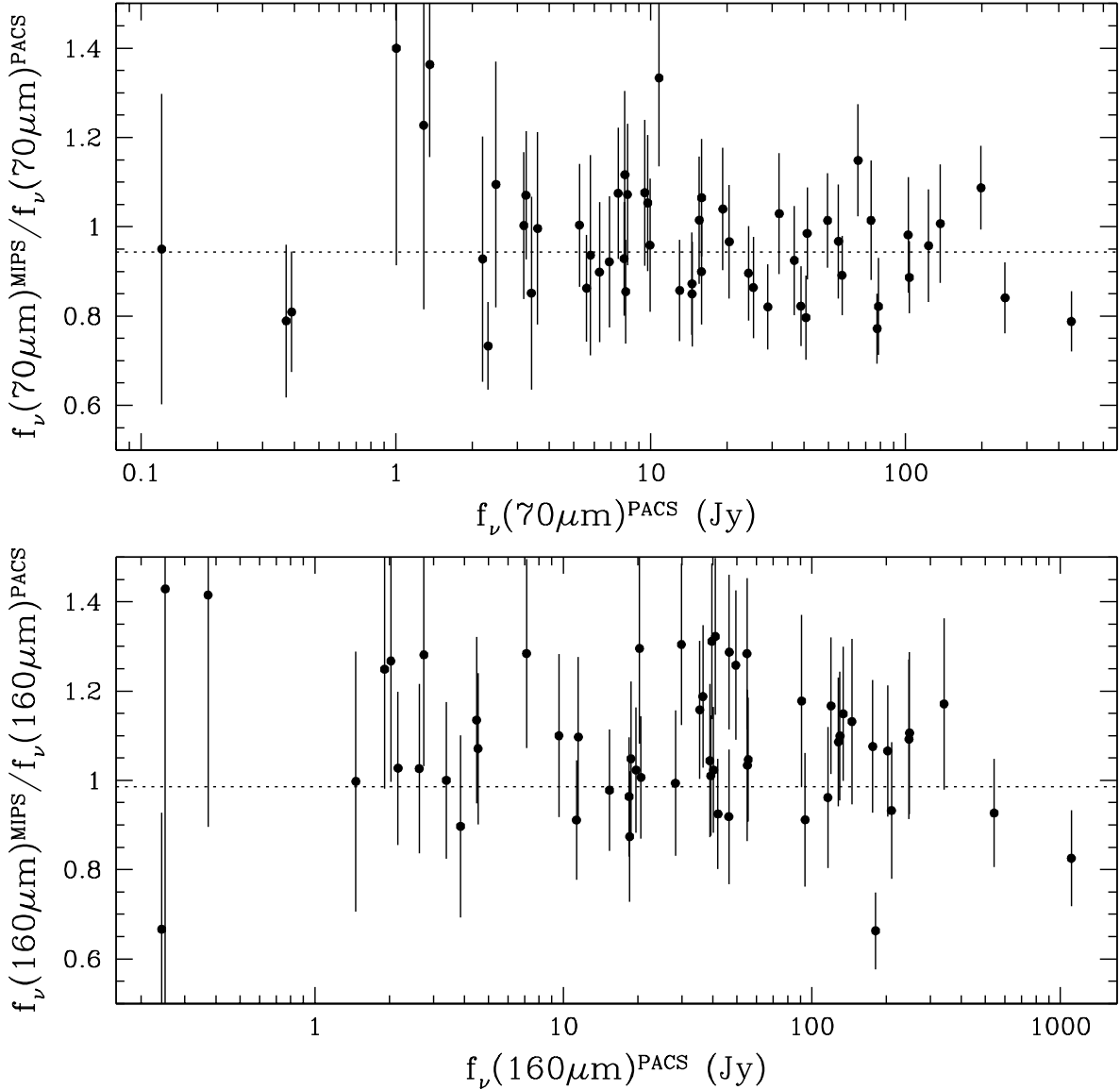}
 \caption{A comparison of spatially-integrated Spitzer/MIPS and Herschel/PACS photometry at 70\m\ (top) and 160\m\ (bottom) for all sample galaxies.  The horizontal dotted lines indicate the ratios corresponding to perfect agreement between data taken by the two observatories, after accounting for differences in the Spitzer and Herschel spectral responses and calibration schemes; for typical galaxy spectral energy distributions, the dotted lines differ from unity by a factor of 1.06 and 1.015 for the 70 and 160\m\ comparisons, respectively.}
 \label{fig:mips.vs.pacs}
\end{figure}

The suite of far-infrared filter bandpasses available for {\it Herschel} and archival {\it Spitzer} data allows a direct comparison of the global flux densities measured for the SINGS and KINGFISH galaxy samples.  The {\it Spitzer} and {\it Herschel} fluxes at 70 and 160\m\ are on average consistent; Figure~\ref{fig:mips.vs.pacs} shows the (error-weighted) ratios of {\it Spitzer}/{\it Herschel} flux densities agree fairly well, after accounting for differences in the {\it Spitzer} and {\it Herschel} spectral responses and calibration schemes.  The agreement at 70\m\ is within 3\% (with a 12\% dispersion in the ratio), while at 160\m\ the MIPS flux densities are typically 6\% larger (with a 16\% dispersion).  Galaxies fainter than $\sim$1~Jy show a much larger dispersion in these ratios, but the flux densities for these targets are more susceptible to errors in sky estimation (\S~\ref{sec:skysub}). 

\begin{figure}
 \plotone{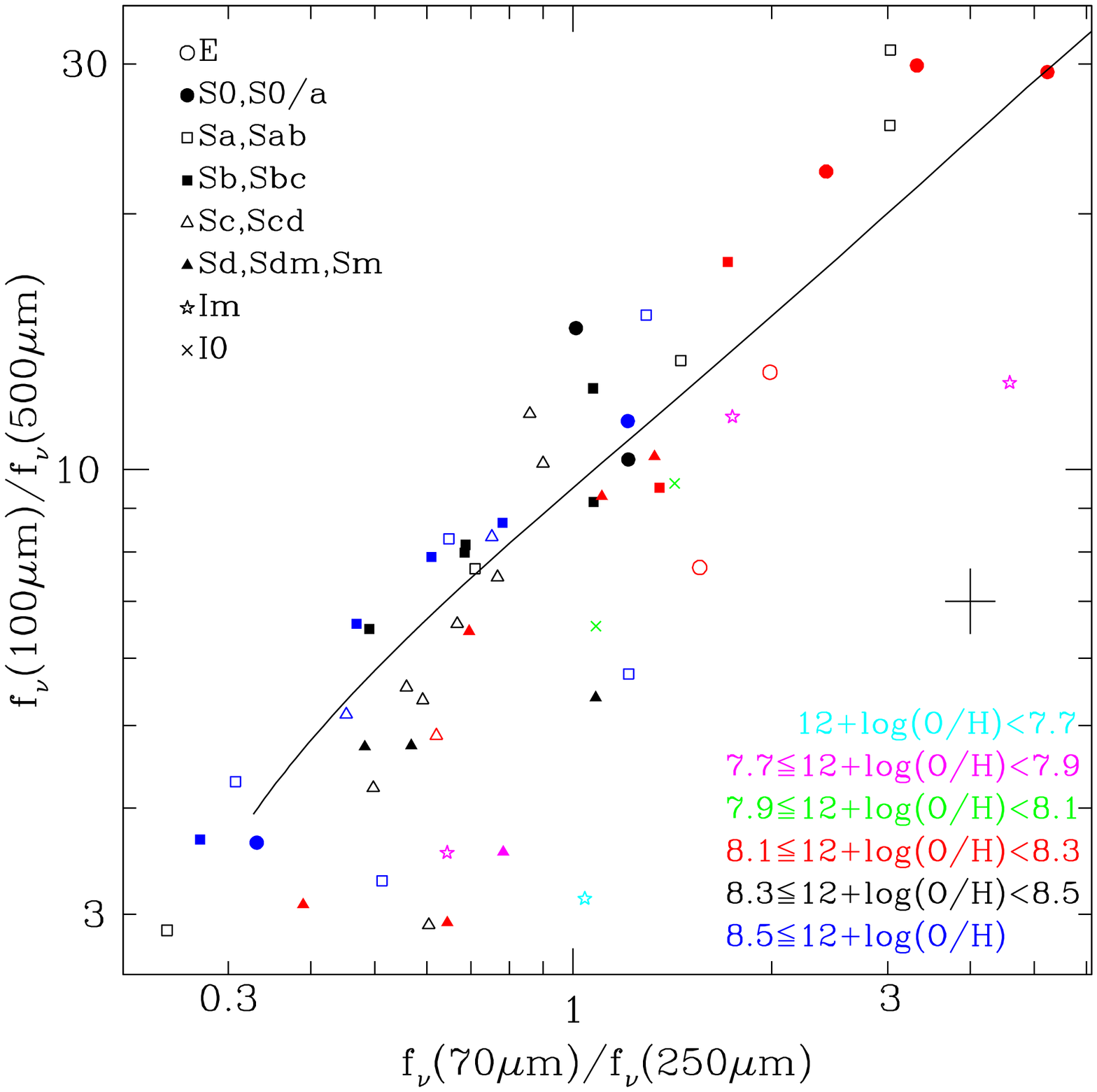}
 \caption{A far-infrared/sub-millimeter color-color diagram for the KINGFISH sample.  ``Characteristic'' oxygen abundances are taken from Table~9 of \cite{moustakas10}, using the \cite{pilyugin05} metallicity scale; if unavailable in Table~9, then the value is computed using the $B$ luminosity and Equation~10 of \cite{moustakas10}, which uses the same metallicity scale.  The solid line indicates the sequence of model spectral energy distributions of \cite{dale02}, derived from the average global trends for a sample of normal star-forming galaxies observed by \ISO\ and \IRAS.  A set of typical error bars is provided.}
 \label{fig:colors}
\end{figure}

\begin{figure}
 \plotone{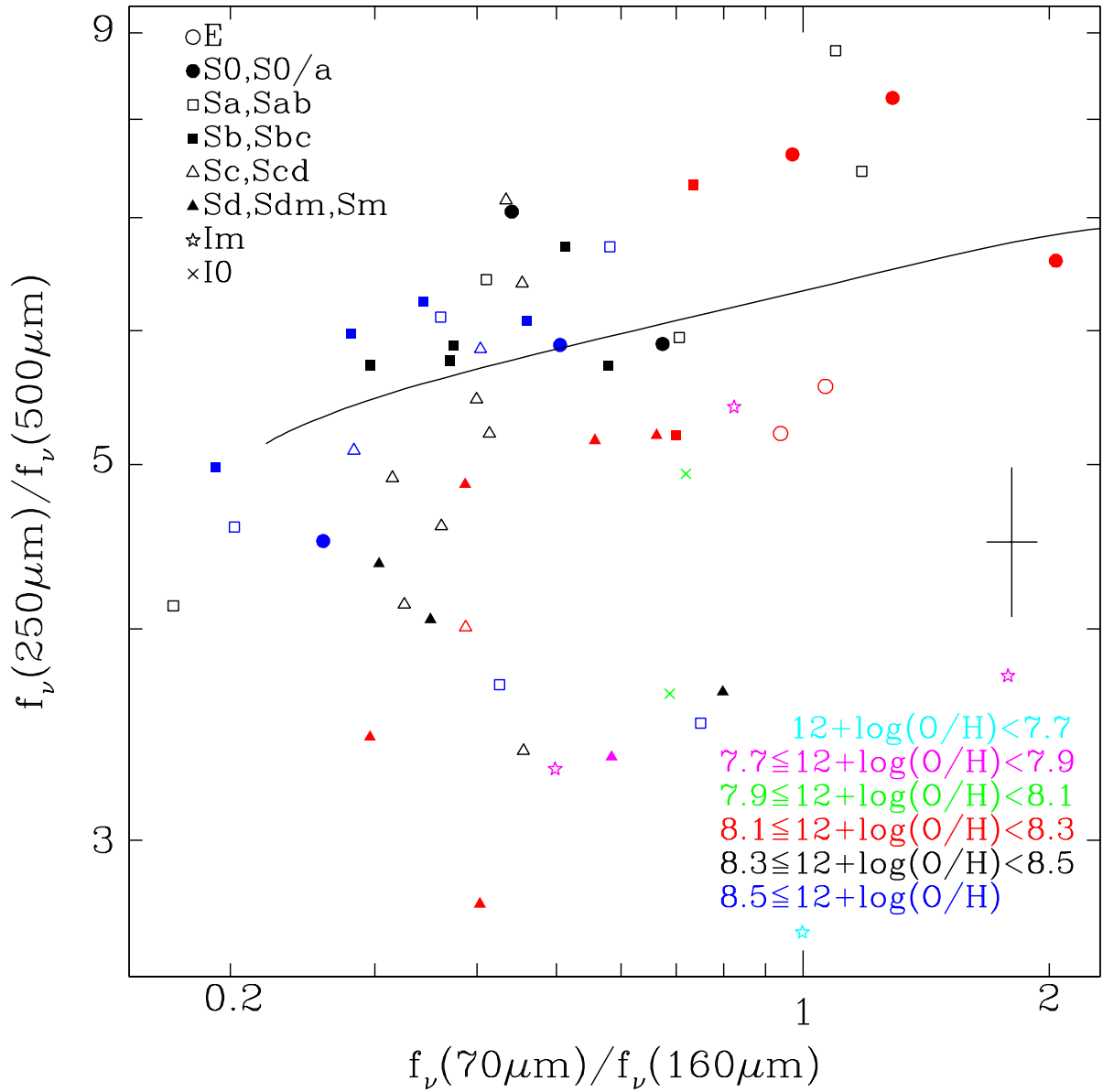}
 \caption{A second far-infrared/sub-millimeter color-color diagram for the KINGFISH sample (see also Figure~\ref{fig:colors}).}
 \label{fig:colors2}
\end{figure}

Figures~\ref{fig:colors} and \ref{fig:colors2} provide a color-color snapshot of the {\it Herschel} global photometry.  As expected, a clear correlation is seen in Figure~\ref{fig:colors} when the flux density ratios on both axes involve wavelengths that straddle the broad infrared peak of emission for most galaxies.  Figure~\ref{fig:colors2}, on the other hand, demonstrates that the galaxy spectral energy distributions in general do not form a simple one-parameter sequence.  Galaxies are more complicated, with mixtures of dust temperatures and distributions of grain properties that vary from one galaxy to another.  The galaxy types are fairly well distributed in terms of their infrared/sub-millimeter colors, though the Sc and Sd spirals tend to cluster toward cooler infrared colors (i.e., smaller values of $f_\nu(70\micron)/f_\nu(160\micron)$, $f_\nu(70\micron)/f_\nu(250\micron)$, and $f_\nu(100\micron)/f_\nu(500\micron)$) and the Magellanic Irregulars (type Im) have relatively large 500\m\ flux densities.  This latter issue will be revisited in \S~\ref{sec:submm}.  

\subsection {The Observed Spectral Energy Distributions} 

\begin{figure}
 \plotone{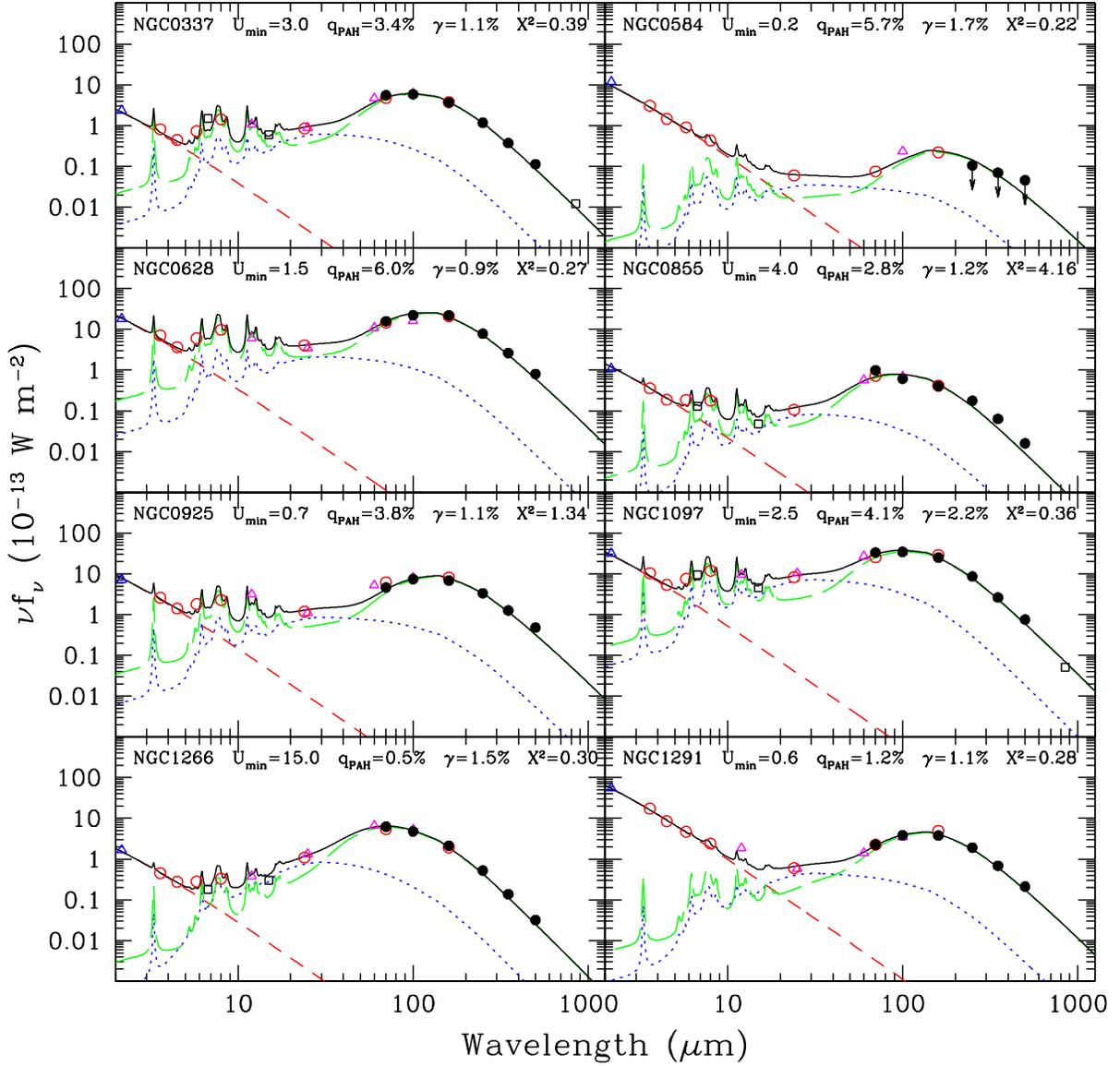}
 \caption{Globally-integrated infrared/sub-millimeter spectral energy distributions for all the galaxies in the KINGFISH sample, sorted by Right Ascension.  Herschel data are represented by filled circles and ancillary data are indicated by open symbols (triangles: {\it 2MASS} and {\it IRAS}; circles: {\it Spitzer}; squares: {\it ISO} and SCUBA).  Arrows indicate 5$\sigma$ upper limits.  The solid curve is the sum of a 5000~K stellar blackbody (short dashed) along with models of dust emission from PDRs (dotted; $U>U_{\rm min}$) and the diffuse interstellar medium (long dashed; $U=U_{\rm min}$).  The fitted parameters from these \cite{draineli07} model fits are listed within each panel along with the reduced $\chi^2$ (see \S~\ref{sec:seds} for details).}
 \label{fig:seds}
\end{figure}

Figure~\ref{fig:seds} shows the observed infrared/sub-millimeter spectral energy distributions for the KINGFISH sample.  Included in each panel, when available, are the 2MASS $K_{\rm s}$, ISO 6.75 and 15\m, {\it Spitzer} 3.6, 4.5, 5.8, 8.0, 24, 70, and 160\m, IRAS 12, 25, 60, and 100\m, {\it Herschel} 70, 100, 160, 250, 350, and 500\m, and SCUBA 450 and 850\m\ band fluxes, derived from this work and \cite{dale07,dale09}.   These data nominally reflect the global emission at each wavelength, but as pointed out in \cite{draine07}, a subset of the SCUBA images suffers from various technical and observational issues.  The data processing for scan mapped SCUBA observations (NGC~4254, NGC~4579, and NGC~6946) removes an unknown contribution from extended emission; the areas mapped by SCUBA for NGC~1097, NGC~4321, and NGC~4736 were small and thus any errors in the large aperture corrections determined by \cite{dale07} for these three systems would have a significant impact on their inferred global fluxes; and the extra-nuclear sub-millimeter emission at 850\m\ is unreliably mapped in NGC~4594 due to contamination by an AGN.  Indeed, for all these special cases except NGC~4736, the SCUBA data appear to fall appreciably below expectations based on extrapolations from the superior {\it Herschel} data.

\subsection {Fits to the Observed Spectral Energy Distributions} 
\label{sec:fits}

To extract physical parameters from the broadband spectral data, the spectral energy distributions were fitted with the models of \cite{draineli07}, models based on mixtures of amorphous silicate and graphitic dust grains that effectively reproduce the average Milky Way extinction curve and are consistent with observations of PAH features and the variety of infrared continua in local galaxies.  \cite{draineli07} use the size distributions of \cite{weingartner01} for dust in the diffuse Milky Way, except for adjustment of the parameters that characterize the PAH size distribution.  The \cite{draineli07} dust models use the far-infrared and sub-millimeter opacities for graphite and amorphous silicate from \cite{lidraine01}.  \cite{lidraine01} used the graphite opacity from \cite{drainelee84}, but made small modifications to the amorphous silicate opacity.  The imaginary part of the amorphous silicate dielectric function $\epsilon_2(\lambda)$ was adjusted in order for the model to better match the average high Galactic latitude dust emission spectrum measured by COBE-FIRAS \citep{wright91,reach95,finkbeiner99}.  The adjustments were modest: $\epsilon_2(\lambda)$ was unchanged for $\lambda < 250~\micron$, and modified by less than 12\% for $250~\micron < \lambda < 1100~\micron$.  With this dielectric function for the amorphous silicate component, the \cite{draineli07} model gives generally good agreement with the observed sub-millimeter emission from the Milky Way diffuse interstellar medium.  Thus the \cite{draineli07} model has in effect been ``tuned'' to reproduce the diffuse emission from the local Milky Way.  While the dust model used here is referenced as coming from \cite{draineli07}, in fact two small changes have been incorporated since that publication: i) there have been some small changes in some of the PAH band parameters, and ii) the graphite dielectric function has been modified to broaden out an opacity peak near 30\m.  These changes are described in \cite{aniano12}.

Building upon an idea put forth by \cite{dale01}, \cite{draineli07} model interstellar dust heating within a galaxy with a $\delta$-function in interstellar radiation field intensity $U$ coupled with a power-law distribution $U_{\rm min} < U < U_{\rm max}$,
\be
dM_{\rm dust}/dU = M_{\rm dust} \left[ (1-\gamma) \delta(U-U_{\rm min}) + \gamma {\alpha-1 \over U_{\rm min}^{1-\alpha} - U_{\rm max}^{1-\alpha}} U^{-\alpha} \right],
\ee
where $U$ is normalized to the local Galactic interstellar radiation field, $dM_{\rm dust}$ is the differential dust mass heated by a range of starlight intensities $[U,U+dU]$, $M_{\rm dust}$ is the total dust mass, and $(1-\gamma)$ is the portion of the dust heated by the diffuse interstellar radiation field defined by $U=U_{\rm min}$.  The minimum and maximum interstellar radiation field intensities span $0.01 < U_{\rm min} < 30$ and $3 < \log U_{\rm max} < 8$.  See \S~5.5 of \cite{dale01} for a physical motivation of the power law distribution in $U$, and Figure~4 of \cite{draineli07} for examples of translating $U$ to dust temperature for different grain sizes.

A sum of three different spectral energy distributions is fit to each galaxy: a blackbody of temperature $T_*=5000$~K, which \cite{smith07} find to be a good approximation to the stellar profile beyond 5\m, along with two dust components.  Following \cite{draine07}, the sum can be expressed as
\be
f_{\nu}^{\rm model} = \Omega_* B_\nu(T_*) + {M_{\rm dust} \over 4 \pi D^2} \left[ (1-\gamma) p_\nu^{(0)}(q_{\rm PAH},U_{\rm min}) + \gamma p_\nu(q_{\rm PAH},U_{\rm min},U_{\rm max},\alpha) \right],
\ee
where $\Omega_*$ is the solid angle subtended by stellar photospheres, $D$ is the distance to the galaxy and $\gamma$ and $(1-\gamma)$ are the fractions of the dust mass heated by the ``power-law'' and ``delta-function'' starlight distributions, respectively.  $p_\nu^{(0)}(q_{\rm PAH},U_{\rm min})$ and $p_\nu(q_{\rm PAH},U_{\rm min},U_{\rm max},\alpha)$ are, respectively, the emitted power per unit frequency per unit dust mass for dust heated by a single starlight intensity $U_{\rm min}$, and dust heated by a power-law distribution of starlight intensities $dM/dU \propto U^{-\alpha}$ extending from $U_{\rm min}$ to $U_{\rm max}$.  The $U=U_{\rm min}$ component may be interpreted as the dust in the general diffuse interstellar medium.  The power-law starlight distribution allows for dust heated by more intense starlight, such as in the intense photodissociation regions (PDRs) in star-forming regions.  For simplicity, emission from dust heated by $U>U_{\rm min}$ will be referred to as the ``PDR'' component, and the emission from dust heated by $U=U_{\rm min}$ will be referred to as the ``diffuse ISM'' component.  Finally, the fractional contribution by total dust mass from PAHs, denoted $q_{\rm PAH}$, varies between 0\% and 12\% with a model grid spacing of 0.1\% in $q_{\rm PAH}$.  

\cite{draine07} find that fits to the (SINGS) global spectral energy distributions of nearby galaxies are insensitive to the minimum radiation field intensity, the maximum radiation field intensity, and the power-law parameter $\alpha$.  We adopt their choice to fix $U_{\rm max}=10^6$ and $\alpha=2$ to minimize the number of free parameters.  \cite{draine07} use a minimum value of 0.7 for $U_{\rm min}$, but we choose to extend this range down to 0.01 due to the availability of SPIRE data longward of 160\m\ and the resulting potential for having detected very cold dust emission.  The free parameters $\Omega_*$, $M_{\rm dust}$, $q_{\rm PAH}$, $U_{\rm min}$, and $\gamma$ are found via $\chi^2$ minimization:
\be
\chi^2 = \sum_b {(f_{\nu,b}^{\rm obs} - f_{\nu,b}^{\rm model})^2 \over (\sigma_{b}^{\rm obs})^2 + (\sigma_{b}^{\rm model})^2} , 
\ee
where $f_{\nu,b}^{\rm model}$ is the model flux density obtained after convolving the model with the $b$ filter bandpass, $\sigma^{\rm obs}_{b}$ is the uncertainty in the observed flux density, and $\sigma^{\rm model}_{b}$ is set to $0.1 f^{\rm model}_{\nu,b}$ to allow for the uncertainty intrinsic to the model.

Figure~\ref{fig:seds} displays the fits of the \cite{draineli07} models to the combined broadband observations from the {\it Spitzer} and {\it Herschel} observatories.  The median reduced chi-squared value is near unity ($\sim 0.7$), and with just a few exceptions the fits are quite reasonable.  The most challenging spectral energy distributions to fit involve spatially variable Milky Way cirrus coupled with a faint target, and thus any errors in determining the value of the local sky has a relatively large impact on the inferred fluxes (i.e., dwarf galaxies such as DDO~053, M81~dwarf~B, and the faint elliptical NGC~0584).

A wealth of information can be extracted from such fits.  Figure~\ref{fig:TIR}, for example, uses these fits to provide a glimpse into how the global spectral energy distributions depend on the star formation rate and total infrared luminosity.  The infrared spectral energy distributions typically peak at shorter wavelengths for KINGFISH galaxies with higher star formation rates and infrared luminosities.  There are exceptions to these generalizations, however, especially for lower luminosity systems.  A full tabulation of the output parameters for the KINGFISH sample will be presented in \cite{aniano12}.  Here we restrict our analysis of the output parameters to i) evaluating the impact of including the {\it Herschel} data in these fits, and ii) comparing the dust masses found through \cite{draineli07} fits with those from single-temperature modified blackbody fits.

\subsection{Spectral Energy Distribution Fit Parameters}
\label{sec:seds}

Figure~\ref{fig:ratios} compares (ratios of) the output parameters $\gamma$, $q_{\rm PAH}$, $U_{\rm min}$, and $M_{\rm dust}$ when the fits are executed with and without the inclusion of {\it Herschel} photometry.  All four parameters are relatively unchanged, on average, when {\it Herschel} broadband data are added to those from {\it Spitzer}.  The largest average deviation in the ratio from unity is seen in the top panel, where the fraction of dust heated by photo-dissociation regions found by using both {\it Spitzer} and {\it Herschel} data is on average ($21\pm4$)\% larger than that using just {\it Spitzer} data.  Interestingly, the largest dispersions in the distributions in Figure~\ref{fig:ratios} are for $U_{\rm min}$ and $M_{\rm dust}$, indicating the importance of Herschel data in assessing these parameters.
In addition, all four parameters show ratio distributions that are fairly evenly distributed about their means, though the scatter shrinks for cooler galaxies.  At first blush it may be surprising that the inclusion of {\it Herschel} far-infrared/sub-millimeter has any impact on a parameter such as $q_{\rm PAH}$ that is sensitive to mid-infrared PAH features, but recall that $q_{\rm PAH}$ is the PAH mass abundance with respect to the total dust mass, and clearly {\it Herschel} photometry has an important role in determining the latter.  Finally, even though $U_{\rm min}$ was allowed to go as low as 0.01, the smallest fitted value using the combined {\it Spitzer} and {\it Herschel} datasets is 0.6, similar to the $U_{\rm min}$ floor advocated by \cite{draine07}.

Figure~\ref{fig:ratiosOxygen} shows the same ratios plotted as a function of oxygen abundance \citep{moustakas10}.  The dependence of KINGFISH dust masses on metallicity are consistent with those found by \cite{galametz11} in a study of 52 galaxies with sub-millimeter data: dust masses computed for metal-rich (metal-poor) galaxies are smaller (larger) when sub-millimeter data are included in the fit.  \cite{galametz11} argue that most metal-rich galaxies have their dust emission peak in the far-infrared beyond 160\m, and that sub-millimeter data are required to fine-tune dust measures for such systems.  KINGFISH galaxies with oxygen abundances $12+\log(O/H)$ greater (less) than 8.1 on the empirical \cite{pilyugin05} metallicity scale are computed to have an average of $0.06\pm0.03$~dex less ($0.28\pm0.09$~dex more) dust mass when sub-millimeter data are used.  This demarcation in metal abundance is similar to that studied by others in quantifying, for example, the relative importance of PAH emission in galaxies \citep[e.g.,][]{hunt05,draine07,smith07,engelbracht08}.  In short, perhaps KINGFISH data show metallicity-dependent dust mass trends similar to those found by \cite{galametz11}, but it would be useful to have more data to confirm any such trend.
 
\subsection{Comparison with Dust Masses from Blackbody Fits}

Galaxy dust masses are typically estimated by fitting a single modified blackbody to a selection of infrared/sub-millimeter continuum fluxes,
\be
M_{\rm dust} = {f_\nu D^2 \over \kappa(\nu_0) B_\nu(T_1)} \left({\nu_0 \over \nu}\right)^{\beta_1} ,
\label{eq:Mdust1}
\ee
or in some cases a superposition of two modified blackbodies of two different dust temperatures, 
\be
M_{\rm dust} = \frac{f_\nu D^2}{\kappa(\nu_0)} \left[x B_\nu(T_1) \left({\nu \over \nu_0}\right)^{\beta_1} + (1-x)B_\nu(T_2) \left({\nu \over \nu_0}\right)^{\beta_2}
\right]^{-1}
\label{eq:Mdust2}
\ee
where $\kappa(\nu_0)$ is the dust 
mass absorption
coefficient at the reference frequency $\nu_0$, $T_1$ and $T_2$ are the modeled dust temperatures, $\beta_1$ and $\beta_2$ are the dust emissivity indexes, and $0<x<1$; some authors choose to fix the dust emissivity index(es) \citep[e.g.,][]{dunne01,kovacs06,pascale09}.
While such approaches provide quick and simple routes to gauging the dust mass, they do not capture the full range of dust temperatures inherent to any galaxy.  However, due to their popularity it is instructive to compare blackbody-based dust masses to those determined from more nuanced models.

Figure~\ref{fig:Mdust} compares the dust masses obtained by using a single-temperature modified blackbody (Equation~\ref{eq:Mdust1}) to those obtained in Section~\ref{sec:seds}.  Both approaches utilize the results of \cite{lidraine01} for dust absorption cross sections, and in particular $\nu_0 = c/250\micron = 1.20$~THz and 
$\kappa(\nu_0) \approx 0.48~{\rm m}^2~{\rm kg}^{-1}$ 
are used for the SPIRE 250\m\ band in determining the (modified) blackbody dust mass.  In addition, both the dust temperature $T_{\rm d}$ and the dust opacity coefficient $\beta$ are allowed to freely vary in the fit for each galaxy; the ranges for the fitted values are quite reasonable given that KINGFISH does not contain any extreme objects: $18 \lesssim T_{\rm d} \lesssim 40$~K and $1.2 \lesssim \beta \lesssim 1.9$ (Figure~\ref{fig:TdustBeta}; see \cite{skibba11} for similar results based on modified blackbody fits to KINGFISH targets).  In order to avoid contributions from stochastically heated dust grains in the computation of the blackbody-based dust masses, the top panel of Figure~\ref{fig:Mdust} shows results when only {\it Herschel} photometric bands from 100\m\ through 500\m\ are included in the fits.  Results are not significantly different when 70\m\ data are included (bottom panel); the median ratio in the top (bottom) panel is 0.53 (0.46).
Figure~\ref{fig:Mdust} indicates that single temperature (modified) blackbody dust masses typically underestimate the values obtained through a \cite{draineli07} formalism by nearly a factor of two ($\sim1.9$), and there is a trend toward larger underestimates for galaxies exhibiting cooler far-infrared colors. 

Similar results are obtained after fixing $\beta$ to either 1.5 or 2.0, except for the situation where the blackbody fits are carried out over the wider 70--500\m\ wavelength baseline for $\beta=2.0$.  In that case the fitted dust temperatures are lower in order to compensate for an overly steep emissivity dependence on wavelength, resulting in larger quantities of dust and only a 25\% underestimate in the dust mass compared to those obtained from \cite{draineli07}, echoing the findings in \cite{magrini11}.  
\cite{dunne01} likewise find a factor of two deficiency for single blackbody-based dust masses, in their case compared to the dust mass derived from two (modified) blackbodies \citep[see also][]{skibba11}.  Figure~\ref{fig:Draine_v_BB} shows a primary reason for the discrepancy: even when limited to $\lambda \geq 100\micron$ photometry, single-temperature blackbody fits overestimate the dust temperature, thus underestimating the dust mass.  The single-temperature model does not account for the contribution of warm dust emitting at shorter wavelengths and the temperatures are driven towards higher values in the attempt to fit both the short and long wavelength far-infrared emission.  This effect is accentuated for galaxies with cooler large dust grains whose emission peaks at longer infrared wavelengths.
A more comprehensive and detailed comparison of various dust mass indicators is studied in \cite{gordon12b}.

\subsection{Sub-millimeter Excess}
\label{sec:submm}

As described in \S~\ref{sec:intro}, several studies of dwarf galaxies show significant excess emission at sub-millimeter wavelengths.  Inspection of Figure~\ref{fig:seds} shows that only a few spiral and elliptical KINGFISH galaxies have 500\m\ fluxes that are noticeably above the fitted model curves (e.g., NGC~3049 and NGC~5474).  However, it is interesting that dwarf/irregular/Magellanic galaxies preferentially show this excess.  There are a total of twelve KINGFISH galaxies of Type~Im (Magellanic irregular), Type~I0 (non-Magellanic irregular) or Type~Sm (Magellanic spiral), and of the nine with detections at 500\m, eight show hints of 500\m\ emission above the \cite{draineli07} model fit (IC~2574, Holmberg~I, Holmberg~II, M81~dwarf~B, NGC~2915, NGC~4236, NGC~5398, and NGC~5408; see also Figure~\ref{fig:colors}).  Quantitatively, the observed 500\m\ excess can be defined with respect to the model prediction at 500\m, namely
\be
\xi(500\mu{\rm m}) = { \nu f_\nu(500\mu{\rm m})_{\rm observed} - \nu f_\nu(500\mu{\rm m})_{\rm model}  \over \nu f_\nu(500\mu{\rm m})_{\rm model} }.
\ee
A dozen KINGFISH galaxies show $\xi(500\mu{\rm m}) > 0.6$, including all eight of the dwarf/irregular/Magellanic galaxies listed above.  However, it should be noted that interpreting $\xi(500\mu{\rm m})$ for these systems is complicated by the fact that they are typically faint in the far-infrared/sub-millimeter \cite[e.g.,][]{walter07} and thus their measured flux values are the least reliable.  Nonetheless, it is interesting that the lowest metallicity objects in the KINGFISH sample are the sources most likely to show a sub-millimeter excess and thus potentially harbor the coldest dust or have peculiar dust grain characteristics.  A detailed analysis of the sub-millimeter excess in KINGFISH galaxies is being carried out by \cite{galametz11}, \cite{gordon12a}, and \cite{hunt12}.

\section {Discussion and Summary}
\label{sec:summary}

Spatially-integrated far-infrared and sub-millimeter flux densities from the {\it Herschel} Space Observatory are provided for the 61 objects in the KINGFISH sample of nearby galaxies.  All but three galaxies are detected in the far-infrared by PACS and all but four galaxies are detected in the sub-millimeter by SPIRE.  The (color-corrected) {\it Herschel} PACS 70\m\ global flux densities agree with {\it Spitzer} MIPS 70\m\ counterparts to within 3\% (with a 12\% dispersion), on average; the MIPS 160\m\ flux densities are typically 6\% larger than the PACS 160\m\ flux densities (with a 16\% dispersion).  

The dust emission models described in \cite{draineli07} and \cite{draine07} are fit to the combined {\it Spitzer} and {\it Herschel} 3.6--500\m\ dataset.  The fits provide constraints on the total dust mass $M_{\rm dust}$, the PAH mass fraction $q_{\rm PAH}$, and the characteristics of the radiation fields that heat the dust including the fraction $\gamma$ of the dust mass that is located in regions with $U>U_{\rm min}$, and the complementary fraction $1-\gamma$ that is located in the general diffuse interstellar medium.  A full tabulation of the fit parameters will be presented in \cite{aniano12}; analysis of the fit results here is restricted to comparisons between fits with and without inclusion of the {\it Herschel} data.  In general, the fits to {\it Spitzer}$+${\it Herschel} data produce parameter values that are consistent, to within a factor of 2, with those when just {\it Spitzer} data are fitted.  However, the KINGFISH galaxies with oxygen abundances less than $12 + \log(O/H) \lesssim 8$ tend to show larger dust masses and smaller PAH mass fractions when SPIRE data are included in the fits.  A similar characteristic oxygen abundance has been noted in other studies of the PAH abundance in galaxies \citep[e.g.,][]{hunt05,draine07,smith07,engelbracht08}.  In addition, 
the fraction of the dust mass located in regions with $U>U_{\rm min}$, $\gamma$, is ($21 \pm 4$)\% larger when Herschel data are included in the fits.  For $\alpha=2$ the fraction of the total dust luminosity contributed by regions with $U>100$ is given by Equation~29 of \cite{draineli07}:
\be
f(L_{\rm dust},U>10^2)=\frac{\gamma \ln(U_{\rm max}/10^2)} {(1-\gamma)(1-U_{\rm min}/U_{\rm max})+\gamma\ln(U_{\rm max}/U_{\rm min})}.
\ee  
This parameter is ($16 \pm 4$)\% larger when Herschel data are included in the fits.  These subtle differences in the fits presumably reflect the unprecedented ability of Herschel to properly account for contributions from cold dust grains, grains that sustain their meager thermal emission through heating by the diffuse radiation field that permeates a galaxy's interstellar medium.  

The presence of an excess of emission in the sub-millimeter has been noted in the literature, particularly for low-metallicity galaxies.  Most KINGFISH galaxies are well modeled by spectral energy distributions consistent with emission curves from the Milky Way and nearby galaxies, without needing to invoke an additional cold dust component.  However, eight of the nine dwarf/irregular/Magellanic galaxies with detections at 500\m\ show evidence for significant excess of emission at this wavelength, at least with respect to the expectations based on the \cite{draineli07} model fits.  These excesses, in fact, are the reason their dust masses are larger when Herschel data are included in the fits described above, assuming these excesses are attributable to increased quantities of very cold dust.  
It is unclear why low-metallicity dwarf irregular galaxies exhibit a propensity for conspicuous cold dust emission. 
In fact, their spectral energy distributions do not typically peak at longer wavelengths than is seen for the more metal rich galaxies; the KINGFISH dwarf galaxies are not colder than average, they simply show 500\m\ excesses.  Perhaps such environments promote unusual dust emissivities that lead to the observed excesses \citep[see also][for additional explanations]{bot10}.

It is commonplace to find in the literature dust masses based on fits to single modified blackbody profiles, with the dust temperature and dust emissivity modifier $\nu^\beta$ serving as potential free parameters.  Blackbody-based dust masses are on average a factor of $\sim1.9$ smaller than those obtained through fits of \cite{draineli07} models, and the disagreement is larger for galaxies with cooler far-infrared colors.   This systematic difference is due to the superior ability of the \cite{draineli07} dust model to represent the dust emission spectrum from the near-infrared through the sub-millimeter, with (for a given value of $q_{\rm PAH}$) a single dust opacity function $\kappa_\nu$, but allowing for a distribution of starlight heating intensities and resulting dust temperature distributions.

\acknowledgements 
{\em Herschel} is an ESA space observatory with science instruments
provided by European-led Principal Investigator consortia and with
important participation from NASA.  IRAF, the Image Reduction and Analysis Facility, has been developed by the National Optical Astronomy Observatories and the Space Telescope Science Institute.  



\begin{deluxetable}{lllccrrrrr}
\tablenum{1}

\def\p{$\pm$}
\tabletypesize{\scriptsize}
\tablecaption{Galaxy Sample}
\tablewidth{0pc}
\tablehead{
\colhead{Galaxy} &
\colhead{Alternative} &
\colhead{Optical} &
\colhead{$E(B-V)$} &
\colhead{$\alpha_0$~\&~$\delta_0$} &
\colhead{$D_{25}$} &
\colhead{2$a$} &
\colhead{2$b$} &
\colhead{PA} &
\colhead{TIR} 
\\
\colhead{} &
\colhead{Name} &
\colhead{Morph.} &
\colhead{(mag.)} &
\colhead{(J2000)} &
\colhead{($^\prime$)} &
\colhead{($\arcsec$)} &
\colhead{($\arcsec$)} &
\colhead{($\degr$)} &
\colhead{($L_\odot$)} 
}
\startdata
NGC0337&        &SBd   &0.112&005950.7$-$073444& 2.9& 253& 194&140&   10.1\\
NGC0584&        &E4    &0.042&013120.6$-$065205& 4.2& 326& 278& 60&    8.8\\
NGC0628&UGC01149&SAc   &0.070&013642.4$+$154711&10.5& 879& 808& 90&    9.9\\
NGC0855&UGC01718&E     &0.071&021403.7$+$275237& 2.6& 259& 169& 60&    8.6\\
NGC0925&UGC01913&SABd  &0.076&022713.6$+$333504&10.5& 735& 486&105&    9.7\\
NGC1097&UGCA041 &SBb   &0.027&024618.0$-$301642& 9.3& 758& 612&130&   10.7\\
NGC1266&        &SB0   &0.098&031600.7$-$022541& 1.5& 234& 232&  0&   10.4\\
NGC1291&        &SB0/a &0.013&031717.9$-$410616& 9.8& 884& 836& 90&    9.5\\
NGC1316&FornaxA &SAB0  &0.021&032241.2$-$371210&12.0& 864& 583& 50&    9.9\\
NGC1377&        &S0    &0.028&033639.0$-$205408& 1.8& 181& 162& 90&   10.1\\
NGC1404&        &E1    &0.011&033852.3$-$353540& 3.3& 524& 369&149&\nodata\\
IC0342 &UGC02847&SABcd &0.558&034648.5$+$680538&21.4&1621&1433&100&   10.1\\
NGC1482&        &SA0   &0.040&035439.0$-$203009& 2.5& 349& 310&119&   10.6\\
NGC1512&        &SBab  &0.011&040355.6$-$432149& 8.9&1001& 928& 83&    9.5\\
NGC2146&UGC03429&Sbab  &0.096&061835.6$+$782129& 6.0& 236& 235&120&   11.0\\
HoII   &UGC04305&Im    &0.032&081910.8$+$704320& 7.9& 554& 465& 60&    7.8\\
DDO053 &UGC04459&Im    &0.038&083407.4$+$661043& 1.5& 155& 142& 90&    7.0\\
NGC2798&UGC04905&SBa   &0.020&091723.1$+$415957& 2.6& 235& 232& 90&   10.6\\
NGC2841&UGC04966&SAb   &0.015&092203.3$+$505837& 8.1& 629& 334&150&   10.1\\
NGC2915&        &I0    &0.275&092609.4$-$763736& 1.9& 183& 132&110&    7.6\\
HoI    &UGC05139&IABm  &0.050&094033.6$+$711120& 3.6& 264& 219& 63&    7.1\\
NGC2976&UGC05221&SAc   &0.071&094715.3$+$675509& 5.9& 541& 353&144&    8.9\\
NGC3049&UGC05325&SBab  &0.038&095449.6$+$091614& 2.2& 218& 160& 29&    9.5\\
NGC3077&UGC05398&I0pec &0.067&100317.5$+$684354& 5.4& 488& 436& 64&    8.9\\
M81dwB &UGC05423&Im    &0.081&100531.2$+$702151& 0.9& 134&  90&139&    6.5\\
NGC3190&UGC05559&SAap  &0.025&101805.7$+$214957& 4.4& 334& 196&117&    9.9\\
NGC3184&UGC05557&SABcd &0.017&101815.6$+$412542& 7.4& 614& 538&169&   10.0\\
NGC3198&UGC05572&SBc   &0.012&101954.8$+$453301& 8.5& 518& 315& 35&   10.0\\
IC2574 &UGC05666&SABm  &0.036&102823.9$+$682505&13.2& 864& 486& 59&    8.3\\
NGC3265&UGC05705&E     &0.024&103106.8$+$284751& 1.3& 184& 175& 50&    9.4\\
NGC3351&M095    &SBb   &0.028&104358.1$+$114210& 7.4& 592& 441& 11&    9.9\\
NGC3521&UGC06150&SABbc &0.057&110548.1$-$000127&11.0& 926& 455&165&   10.5\\
NGC3621&UGCA232 &SAd   &0.081&111818.3$-$324855&12.3& 791& 555&160&    9.9\\
NGC3627&M066    &SABb  &0.033&112013.4$+$125927& 9.1& 745& 486&167&   10.4\\
NGC3773&UGC06605&SA0   &0.027&113813.1$+$120644& 1.2& 118& 116&  0&    8.8\\
NGC3938&UGC06856&SAc   &0.021&115250.3$+$440715& 5.4& 504& 468&  0&   10.3\\
NGC4236&UGC07306&SBdm  &0.015&121643.2$+$692719&21.9&1240& 369&162&    8.7\\
NGC4254&M099    &SAc   &0.039&121849.7$+$142519& 5.4& 519& 420& 60&   10.6\\
NGC4321&M100    &SABbc &0.026&122254.8$+$154907& 7.4& 558& 483& 40&   10.5\\
NGC4536&UGC07732&SABbc &0.018&123427.5$+$021113& 7.6& 454& 376&120&   10.3\\
NGC4559&UGC07766&SABcd &0.018&123558.1$+$275752&10.7& 576& 327&140&    9.5\\
NGC4569&M090    &SABab &0.047&123650.2$+$131001& 9.5& 593& 327& 21&    9.7\\
NGC4579&M058    &SABb  &0.041&123743.8$+$114858& 5.9& 325& 271& 90&   10.1\\
NGC4594&M104    &SAa   &0.051&123959.6$-$113726& 8.7& 767& 669& 90&    9.6\\
NGC4625&UGC07861&SABmp &0.018&124154.8$+$411623& 2.2& 298& 214&100&    8.8\\
NGC4631&UGC07865&SBd   &0.017&124204.2$+$323219&15.5& 901& 240& 85&   10.4\\
NGC4725&UGC07989&SABab &0.012&125027.7$+$252948&10.7& 689& 523& 30&    9.9\\
NGC4736&M094    &SAab  &0.018&125055.2$+$410652&11.2& 944& 899&  0&    9.8\\
DDO154 &UGC08024&IBm   &0.009&125407.6$+$270916& 3.0& 216& 126& 50&\nodata\\
NGC4826&M064    &SAab  &0.041&125643.3$+$214048&10.0& 716& 427&114&    9.6\\
DDO165 &UGC08201&Im    &0.024&130625.9$+$674229& 3.5& 263& 161& 90&\nodata\\
NGC5055&M063    &SAbc  &0.018&131549.2$+$420147&12.6&1097& 711& 80&   10.3\\
NGC5398&Tololo89&SBdm  &0.066&140121.2$-$330402& 2.8& 198& 146&  0&    8.6\\
NGC5457&M101    &SABcd &0.009&140325.0$+$542429&28.8&1800&1446& 37&   10.4\\
NGC5408&        &IBm   &0.068&140321.1$-$412241& 1.6& 256& 209& 67&    8.3\\
NGC5474&UGC09013&SAcd  &0.011&140500.8$+$533920& 4.8& 412& 373& 90&    8.7\\
NGC5713&UGC09451&SABbcp&0.039&144011.4$-$001726& 2.8& 225& 225& 90&   10.5\\
NGC5866&UGC09723&S0    &0.013&150628.8$+$554551& 4.7& 500& 306&129&    9.8\\
NGC6946&UGC11597&SABcd &0.342&203449.2$+$600959&11.5& 953& 928&  0&   10.5\\
NGC7331&UGC12113&SAb   &0.091&223704.3$+$342435&10.5& 683& 335&168&   10.7\\
NGC7793&        &SAd   &0.019&235749.9$-$323525& 9.3& 716& 526& 98&    9.3\\

\enddata
\tablecomments{\footnotesize $D_{25}$ is the diameter of the $B$ band isophote defined at 25~mag~arcsec$^{-2}$.  $2a$ and $2b$ are the lengths of the major and minor axes used in the elliptical aperture photometry described herein; the position angle of the aperture's major axis is measured east of north.  The total infrared listed in the last column is derived from Equation~4 of \cite{dale02} and the far-infrared fluxes in \cite{dale07,dale09} and from \cite{engelbracht08} for IC~0342 and NGC~2146.}
\label{tab:sample}
\end{deluxetable}

\clearpage
\begin{deluxetable}{lrrrrrr}
\tablenum{2}
\def\a{\tablenotemark{a}}
\def\p{$\pm$}
\tabletypesize{\scriptsize}
\tablecaption{Far-Infrared/Sub-millimeter Flux Densities}
\tablewidth{0pc}
\tablehead{
\colhead{Galaxy} &
\colhead{PACS} &
\colhead{PACS} &
\colhead{PACS} &
\colhead{SPIRE} &
\colhead{SPIRE} &
\colhead{SPIRE} 
\\
\colhead{} &
\colhead{70\m} &
\colhead{100\m} &
\colhead{160\m} &
\colhead{250\m} &
\colhead{350m} &
\colhead{500\m} 
\\
\colhead{} & 
\colhead{(Jy)} &
\colhead{(Jy)} &
\colhead{(Jy)} &
\colhead{(Jy)} &
\colhead{(Jy)} &
\colhead{(Jy)} 
}
\startdata
NGC0337  &1.30\p0.07E+1&1.95\p0.10E+1&1.96\p0.10E+1&9.79\p0.70E+0&4.37\p0.31E+0&1.88\p0.14E+0\\
NGC0584  &\nodata      &\nodata      &\nodata      &$<$8.84E-1   &$<$8.19E-1   &$<$7.63E-1   \\
NGC0628  &3.67\p0.18E+1&7.40\p0.37E+1&1.16\p0.06E+2&6.55\p0.47E+1&3.06\p0.22E+1&1.33\p0.10E+1\\
NGC0855  &2.30\p0.12E+0&2.04\p0.12E+0&2.16\p0.12E+0&1.48\p0.11E+0&7.47\p0.65E-1&2.66\p0.40E-1\\
NGC0925  &1.08\p0.06E+1&2.47\p0.12E+1&3.65\p0.18E+1&2.77\p0.20E+1&1.48\p0.11E+1&8.03\p0.58E+0\\
NGC1097  &7.75\p0.39E+1&1.16\p0.06E+2&1.34\p0.07E+2&7.22\p0.51E+1&3.08\p0.22E+1&1.26\p0.09E+1\\
NGC1266  &1.45\p0.07E+1&1.59\p0.08E+1&1.13\p0.06E+1&4.38\p0.31E+0&1.60\p0.12E+0&5.32\p0.55E-1\\
NGC1291  &5.26\p0.32E+0&1.28\p0.07E+1&2.03\p0.11E+1&1.59\p0.11E+1&7.98\p0.59E+0&3.52\p0.29E+0\\
NGC1316  &5.81\p0.33E+0&9.30\p0.50E+0&1.15\p0.06E+1&4.80\p0.37E+0&2.06\p0.19E+0&8.16\p1.33E-1\\
NGC1377  &6.89\p0.35E+0&5.85\p0.30E+0&3.38\p0.19E+0&1.32\p0.10E+0&4.94\p0.47E-1&1.99\p0.32E-1\\
NGC1404  &$<$6.71E-1   &$<$7.06E-1   &$<$1.02E 0   &$<$4.30E-1   &$<$3.99E-1   &$<$3.72E-1   \\
IC0342   &4.48\p0.22E+2&8.47\p0.42E+2&1.11\p0.06E+3&5.95\p0.42E+2&2.61\p0.19E+2&1.02\p0.07E+2\\
NGC1482  &4.07\p0.20E+1&4.95\p0.25E+1&4.20\p0.21E+1&1.68\p0.12E+1&6.35\p0.45E+0&2.21\p0.17E+0\\
NGC1512  &7.99\p0.47E+0&1.38\p0.07E+1&1.87\p0.10E+1&1.56\p0.11E+1&8.66\p0.64E+0&4.20\p0.34E+0\\
NGC2146  &1.98\p0.10E+2&2.32\p0.12E+2&1.81\p0.09E+2&6.55\p0.47E+1&2.33\p0.17E+1&7.45\p0.53E+0\\
HoII     &3.18\p0.35E+0&3.89\p0.45E+0&3.86\p0.63E+0&1.82\p0.16E+0&8.04\p1.08E-1&3.37\p1.69E-1\\
DDO053   &3.90\p0.42E-1&4.80\p1.21E-1&2.50\p1.77E-1&1.86\p0.32E-1&9.99\p2.81E-2&$<$1.25E-1   \\
NGC2798  &2.42\p0.12E+1&2.73\p0.14E+1&2.06\p0.10E+1&8.02\p0.57E+0&2.90\p0.21E+0&1.08\p0.09E+0\\
NGC2841  &9.49\p0.49E+0&2.57\p0.13E+1&4.95\p0.25E+1&3.49\p0.25E+1&1.60\p0.11E+1&7.01\p0.50E+0\\
NGC2915  &1.01\p0.06E+0&1.66\p0.09E+0&1.46\p0.11E+0&9.28\p0.73E-1&5.28\p0.47E-1&2.54\p0.32E-1\\
HoI      &3.71\p0.62E-1&4.21\p0.70E-1&3.72\p1.20E-1&3.56\p0.53E-1&2.23\p0.47E-1&1.35\p0.41E-1\\
NGC2976  &1.92\p0.10E+1&3.58\p0.18E+1&4.64\p0.23E+1&2.50\p0.18E+1&1.17\p0.08E+1&4.79\p0.35E+0\\
NGC3049  &3.40\p0.18E+0&4.59\p0.23E+0&4.54\p0.24E+0&2.80\p0.20E+0&1.41\p0.11E+0&7.97\p0.65E-1\\
NGC3077  &2.04\p0.10E+1&2.79\p0.14E+1&2.83\p0.14E+1&1.43\p0.10E+1&6.47\p0.47E+0&2.89\p0.22E+0\\
M81dwB   &1.21\p0.41E-1&2.01\p0.31E-1&2.42\p0.82E-1&1.87\p0.25E-1&1.03\p0.22E-1&5.66\p2.83E-2\\
NGC3190  &6.30\p0.33E+0&1.06\p0.05E+1&1.54\p0.08E+1&8.88\p0.63E+0&3.71\p0.27E+0&1.38\p0.11E+0\\
NGC3184  &1.55\p0.08E+1&3.47\p0.17E+1&5.49\p0.28E+1&3.43\p0.24E+1&1.53\p0.11E+1&6.73\p0.49E+0\\
NGC3198  &9.75\p0.51E+0&2.00\p0.10E+1&2.99\p0.15E+1&1.96\p0.14E+1&9.95\p0.71E+0&4.74\p0.34E+0\\
IC2574   &5.61\p0.37E+0&7.57\p0.42E+0&9.61\p0.53E+0&7.16\p0.52E+0&4.83\p0.36E+0&2.13\p0.19E+0\\
NGC3265  &2.47\p0.13E+0&3.10\p0.16E+0&2.63\p0.15E+0&1.24\p0.10E+0&5.51\p0.51E-1&2.38\p0.35E-1\\
NGC3351  &2.53\p0.13E+1&4.61\p0.23E+1&5.51\p0.28E+1&3.24\p0.23E+1&1.37\p0.10E+1&5.32\p0.39E+0\\
NGC3521  &7.85\p0.39E+1&1.58\p0.08E+2&2.10\p0.10E+2&1.14\p0.08E+2&4.72\p0.34E+1&1.94\p0.14E+1\\
NGC3621  &4.95\p0.25E+1&9.44\p0.47E+1&1.28\p0.06E+2&7.12\p0.51E+1&3.17\p0.23E+1&1.46\p0.10E+1\\
NGC3627  &1.04\p0.05E+2&1.79\p0.09E+2&2.02\p0.10E+2&9.67\p0.69E+1&3.76\p0.27E+1&1.44\p0.10E+1\\
NGC3773  &1.29\p0.08E+0&1.85\p0.11E+0&1.91\p0.14E+0&1.06\p0.08E+0&4.34\p0.38E-1&1.80\p0.24E-1\\
NGC3938  &1.58\p0.08E+1&2.86\p0.15E+1&3.96\p0.20E+1&2.37\p0.17E+1&1.03\p0.07E+1&4.34\p0.32E+0\\
NGC4236  &7.46\p0.46E+0&1.23\p0.07E+1&1.85\p0.11E+1&1.16\p0.08E+1&7.37\p0.54E+0&4.21\p0.32E+0\\
NGC4254\a&5.64\p0.28E+1&1.06\p0.05E+2&1.30\p0.07E+2&6.57\p0.47E+1&2.66\p0.19E+1&9.16\p0.66E+0\\
NGC4321\a&4.12\p0.21E+1&8.55\p0.43E+1&1.20\p0.06E+2&6.76\p0.48E+1&2.79\p0.20E+1&1.08\p0.08E+1\\
NGC4536\a&3.89\p0.20E+1&5.26\p0.26E+1&5.55\p0.28E+1&2.88\p0.20E+1&1.26\p0.09E+1&5.53\p0.40E+0\\
NGC4559  &1.59\p0.08E+1&3.10\p0.16E+1&4.10\p0.21E+1&2.55\p0.18E+1&1.28\p0.09E+1&6.37\p0.46E+0\\
NGC4569\a&1.46\p0.07E+1&3.04\p0.15E+1&4.03\p0.20E+1&2.24\p0.16E+1&9.41\p0.67E+0&3.67\p0.27E+0\\
NGC4579\a&9.94\p0.51E+0&2.33\p0.12E+1&3.54\p0.18E+1&2.12\p0.15E+1&8.87\p0.63E+0&3.54\p0.26E+0\\
NGC4594  &7.87\p0.49E+0&2.39\p0.13E+1&3.89\p0.20E+1&2.56\p0.18E+1&1.21\p0.09E+1&5.56\p0.41E+0\\
NGC4625  &1.36\p0.12E+0&3.04\p0.20E+0&4.48\p0.23E+0&2.81\p0.21E+0&1.40\p0.11E+0&6.44\p0.62E-1\\
NGC4631  &1.37\p0.07E+2&2.23\p0.11E+2&2.46\p0.12E+2&1.24\p0.09E+2&5.45\p0.39E+1&2.40\p0.17E+1\\
NGC4725\a&7.93\p0.46E+0&2.28\p0.12E+1&4.66\p0.23E+1&3.27\p0.23E+1&1.66\p0.12E+1&7.93\p0.57E+0\\
NGC4736  &1.03\p0.05E+2&1.59\p0.08E+2&1.45\p0.07E+2&7.04\p0.50E+1&2.80\p0.20E+1&1.18\p0.09E+1\\
DDO154   &$<$3.31E-1   &$<$4.27E-1   &$<$2.67E-1   &$<$1.62E-1   &$<$1.50E-1   &$<$1.39E-1   \\
NGC4826  &5.47\p0.27E+1&9.57\p0.48E+1&9.41\p0.47E+1&4.24\p0.30E+1&1.64\p0.12E+1&6.30\p0.46E+0\\
DDO165   &$<$4.13E-1   &$<$5.33E-1   &$<$3.33E-1   &$<$2.01E-1   &$<$1.87E-1   &$<$1.74E-1   \\
NGC5055  &7.34\p0.37E+1&1.70\p0.08E+2&2.48\p0.12E+2&1.50\p0.11E+2&6.42\p0.46E+1&2.61\p0.19E+1\\
NGC5398  &2.19\p0.12E+0&2.98\p0.16E+0&2.75\p0.16E+0&2.03\p0.15E+0&1.05\p0.08E+0&5.52\p0.49E-1\\
NGC5457  &1.23\p0.06E+2&2.43\p0.12E+2&3.41\p0.17E+2&2.08\p0.15E+2&9.69\p0.69E+1&4.53\p0.32E+1\\
NGC5408  &3.60\p0.19E+0&2.65\p0.15E+0&2.02\p0.11E+0&7.85\p0.72E-1&3.86\p0.50E-1&2.09\p0.42E-1\\
NGC5474  &3.24\p0.18E+0&4.61\p0.25E+0&7.12\p0.37E+0&5.37\p0.39E+0&2.91\p0.22E+0&1.58\p0.13E+0\\
NGC5713  &2.89\p0.14E+1&4.03\p0.20E+1&3.93\p0.20E+1&1.68\p0.12E+1&6.39\p0.46E+0&2.30\p0.17E+0\\
NGC5866  &8.12\p0.42E+0&1.67\p0.09E+1&1.84\p0.10E+1&8.04\p0.58E+0&3.14\p0.23E+0&1.14\p0.10E+0\\
NGC6946  &2.46\p0.12E+2&4.35\p0.22E+2&5.42\p0.27E+2&2.74\p0.19E+2&1.09\p0.08E+2&4.28\p0.30E+1\\
NGC7331  &6.53\p0.33E+1&1.32\p0.07E+2&1.76\p0.09E+2&9.53\p0.68E+1&4.06\p0.29E+1&1.65\p0.12E+1\\
NGC7793  &3.20\p0.16E+1&6.58\p0.33E+1&9.11\p0.46E+1&5.63\p0.40E+1&2.84\p0.20E+1&1.39\p0.10E+1\\

\enddata     
\tablecomments{\footnotesize The compact table entry format T.UV$\pm$W.XYEZ implies (T.UV$\pm$W.XY)$\times10^{\rm Z}$.  See \S~\ref{sec:data} for corrections that have been applied to the data.  The uncertainties include both statistical and systematic effects.  5$\sigma$ upper limits are provided for non-detections.  No color corrections have been applied.}
\tablenotetext{a}{\footnotesize SPIRE imaging taken from the Herschel Reference Survey \citep{boselli10}.}
\label{tab:data}
\end{deluxetable}


\addtocounter{figure}{-1}
\begin{figure} 
 \plotone{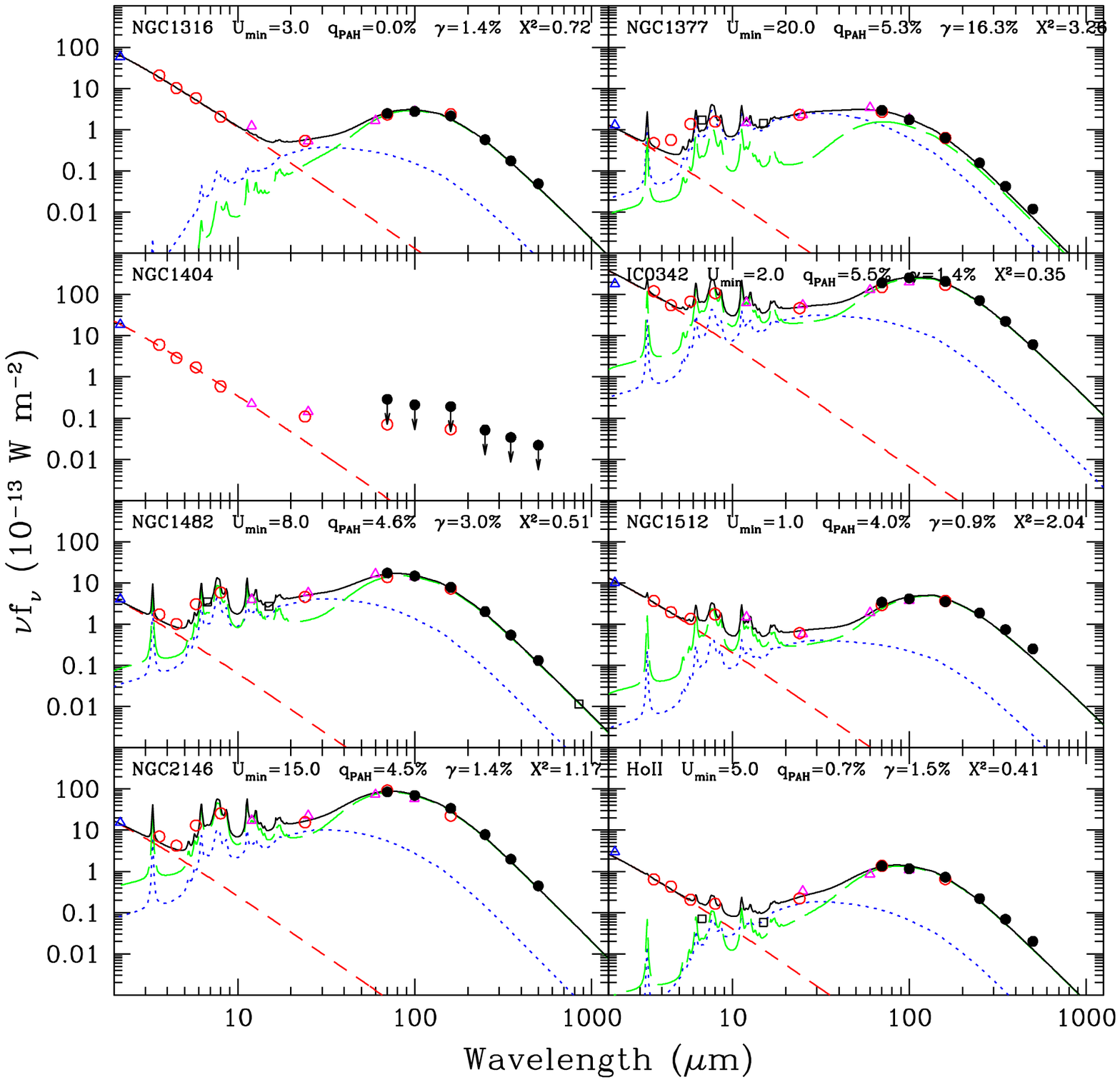} 
 \caption{Globally-integrated infrared/sub-millimeter spectral energy distributions for the KINGFISH sample (continued).}
\end{figure}

\addtocounter{figure}{-1}
\begin{figure} 
 \plotone{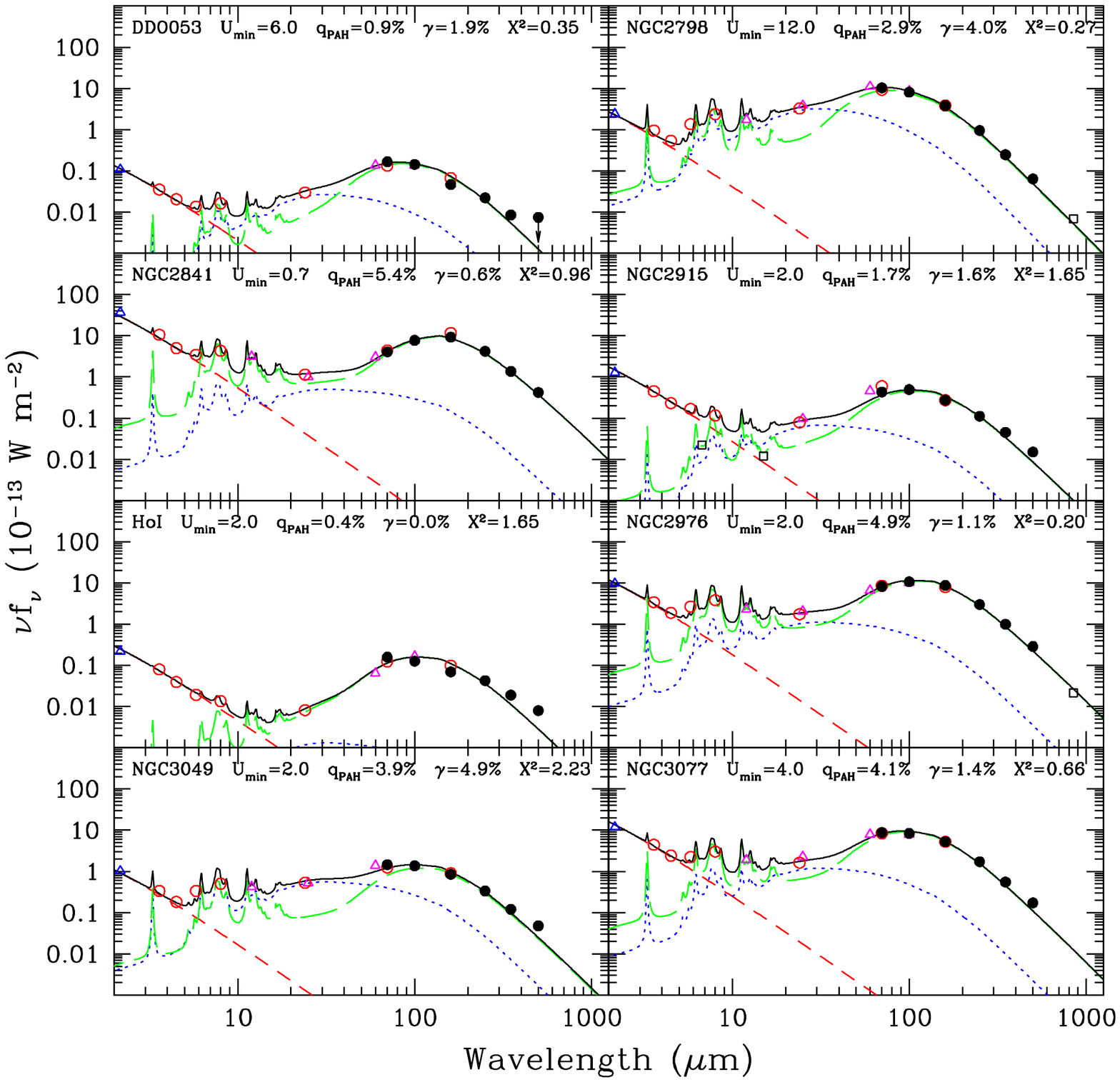} 
 \caption{Globally-integrated infrared/sub-millimeter spectral energy distributions for the KINGFISH sample (continued).}
\end{figure}

\addtocounter{figure}{-1}
\begin{figure} 
 \plotone{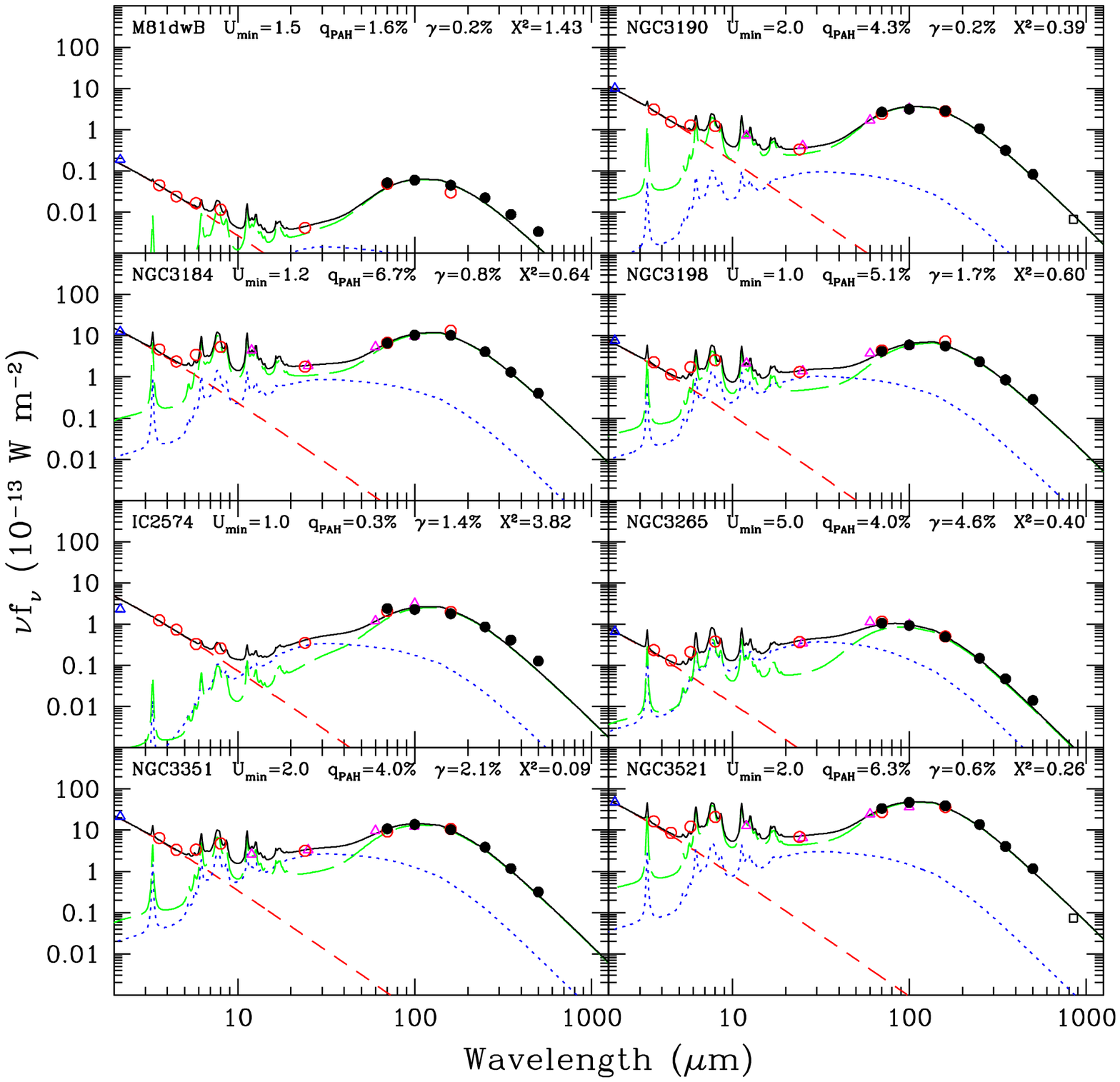} 
 \caption{Globally-integrated infrared/sub-millimeter spectral energy distributions for the KINGFISH sample (continued).}
\end{figure}

\addtocounter{figure}{-1}
\begin{figure} 
 \plotone{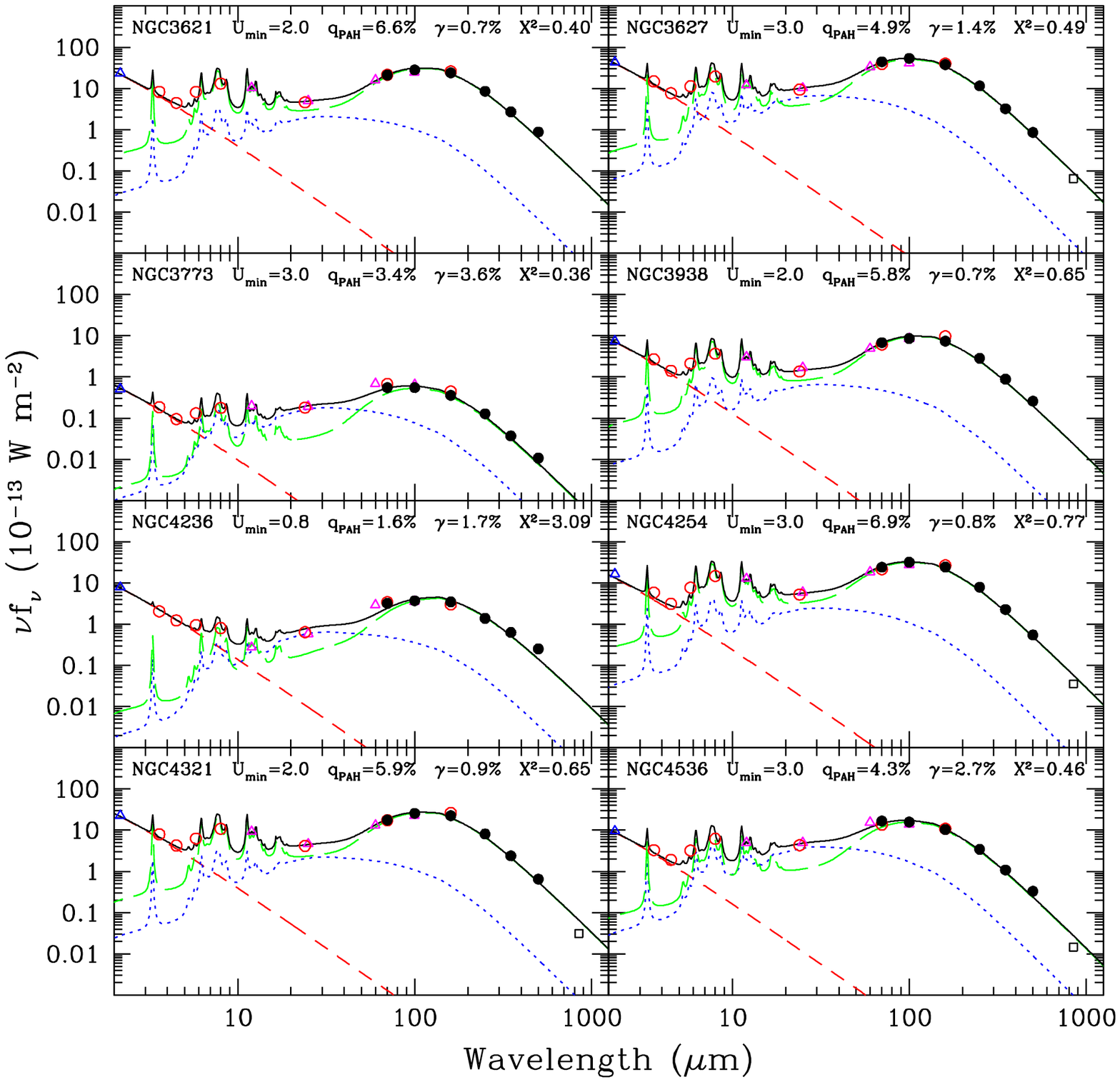} 
 \caption{Globally-integrated infrared/sub-millimeter spectral energy distributions for the KINGFISH sample (continued).}
\end{figure}

\addtocounter{figure}{-1}
\begin{figure} 
 \plotone{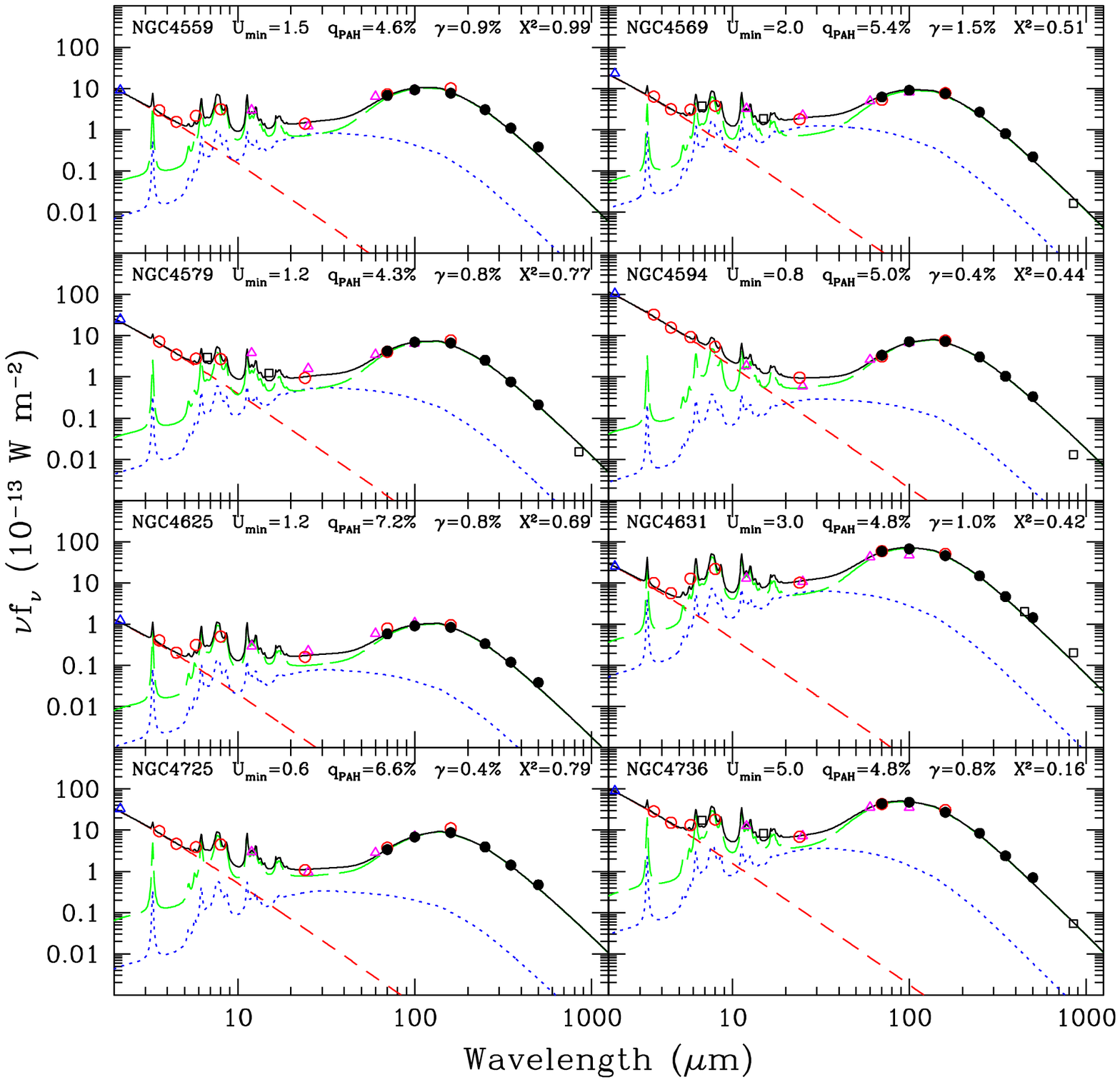} 
 \caption{Globally-integrated infrared/sub-millimeter spectral energy distributions for the KINGFISH sample (continued).}
\end{figure}

\addtocounter{figure}{-1}
\begin{figure} 
 \plotone{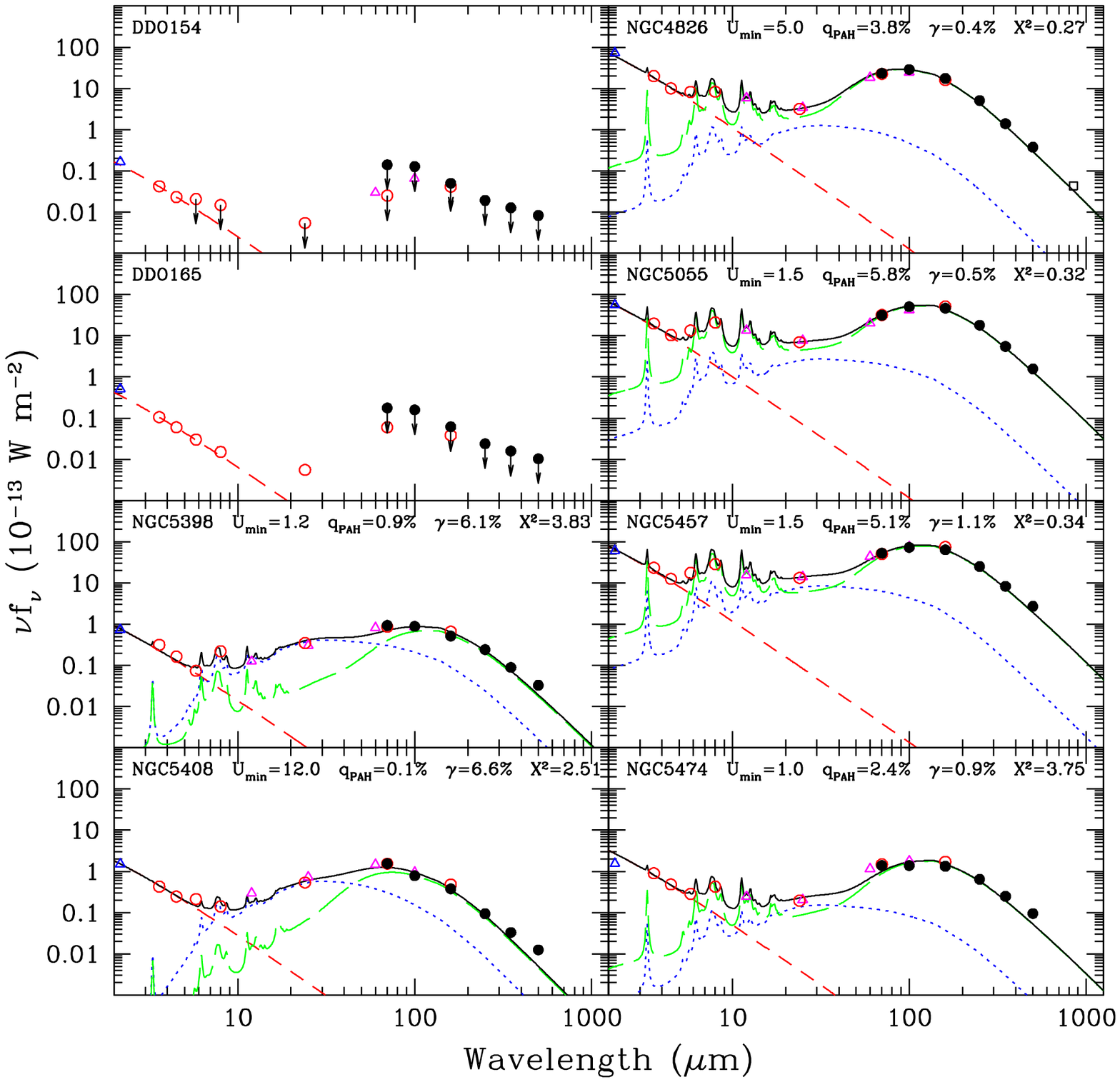} 
 \caption{Globally-integrated infrared/sub-millimeter spectral energy distributions for the KINGFISH sample (continued).}
\end{figure}

\addtocounter{figure}{-1}
\begin{figure} 
 \plotone{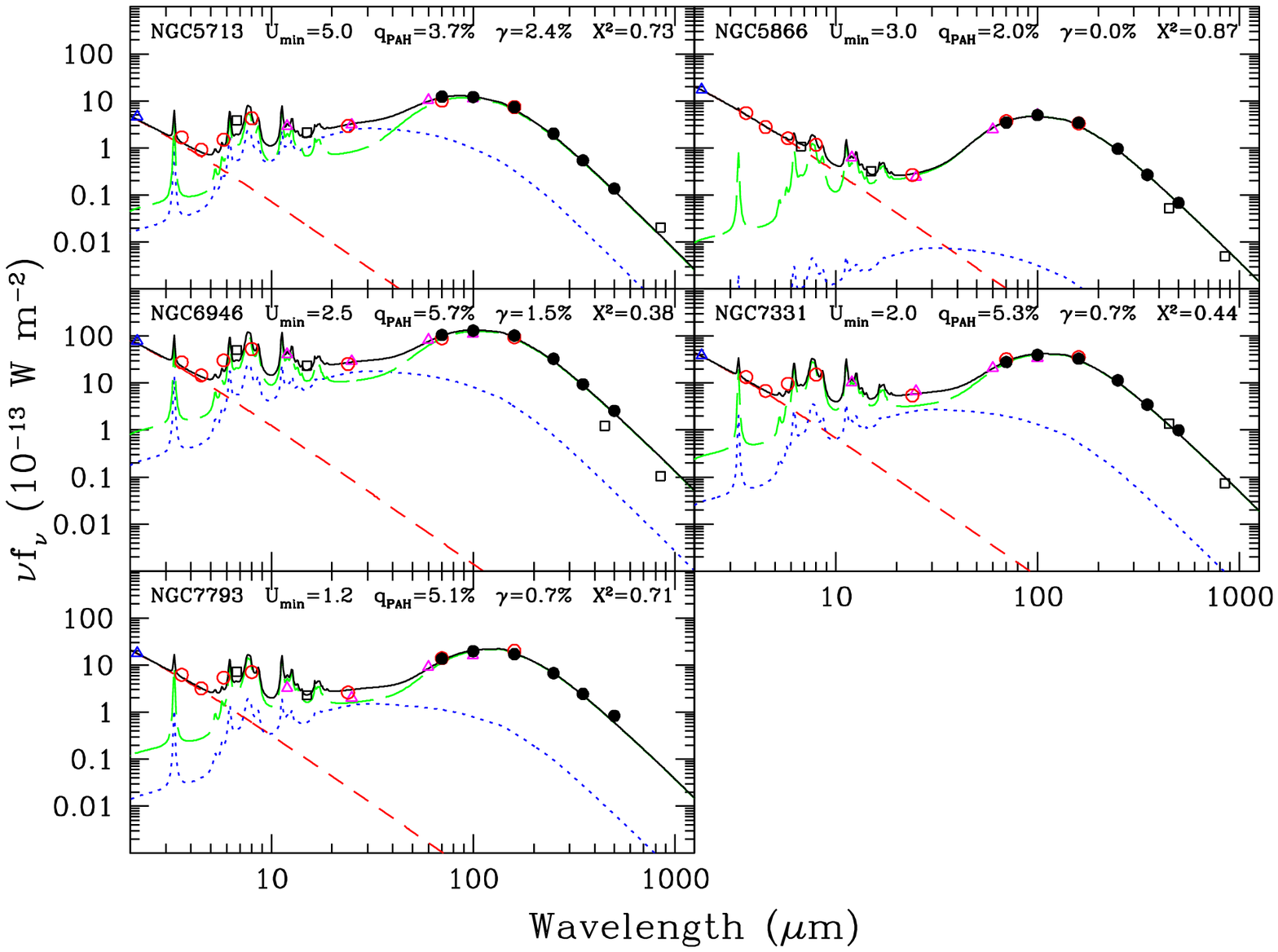} 
 \caption{Globally-integrated infrared/sub-millimeter spectral energy distributions for the KINGFISH sample (continued).}
\end{figure}

\clearpage
\begin{figure} 
 \plotone{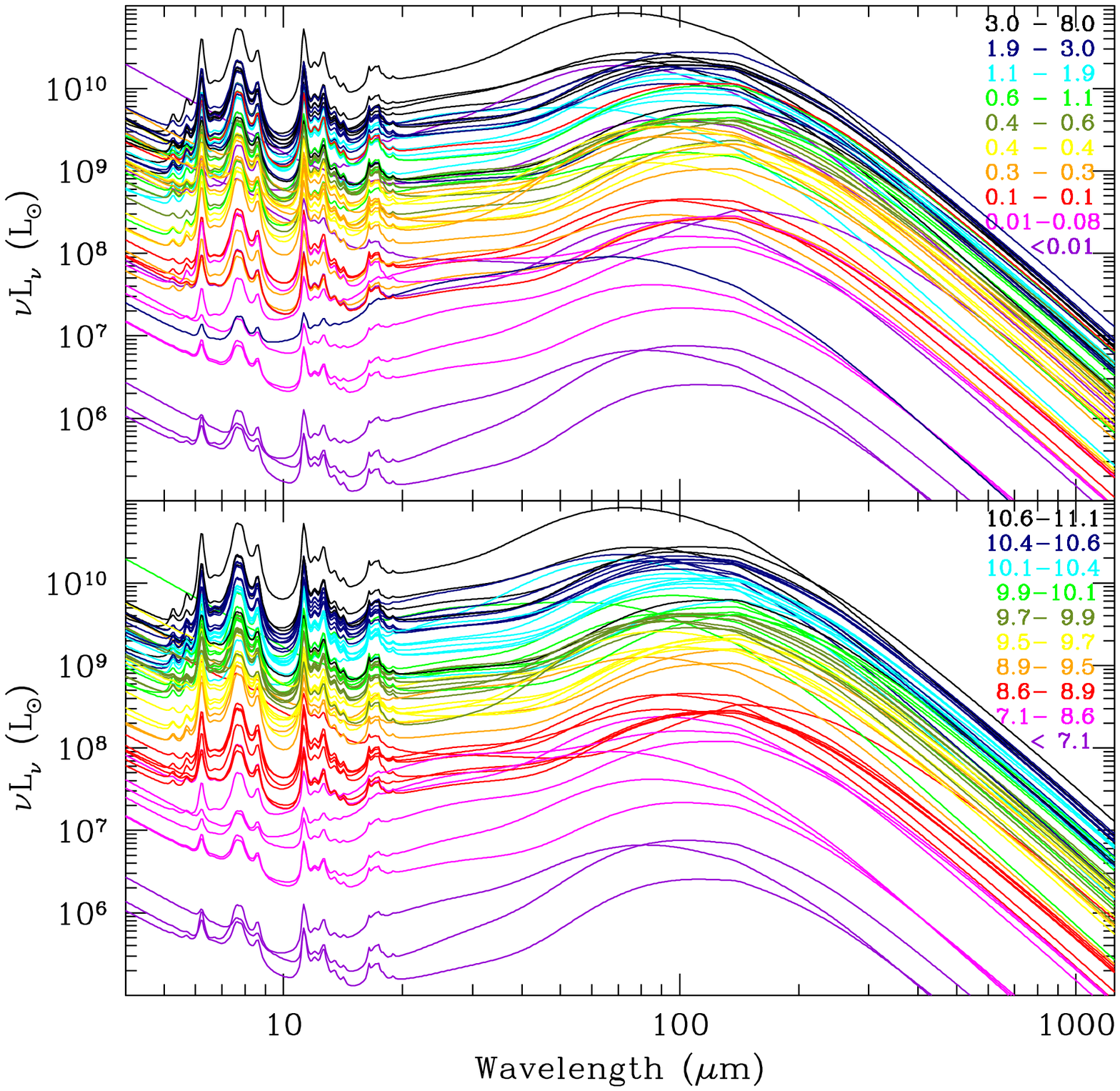} 
 \caption{The collection of \cite{draineli07} model fits, color coded according to star formation rate in $M_\odot~{\rm yr}^{-1}$ in the top panel \citep{kennicutt12} and $\log_{10} (L_{\rm TIR}/L_\odot)$ in the bottom panel (see Table~\ref{tab:sample}).}
 \label{fig:TIR}
\end{figure}

\begin{figure} 
 \plotone{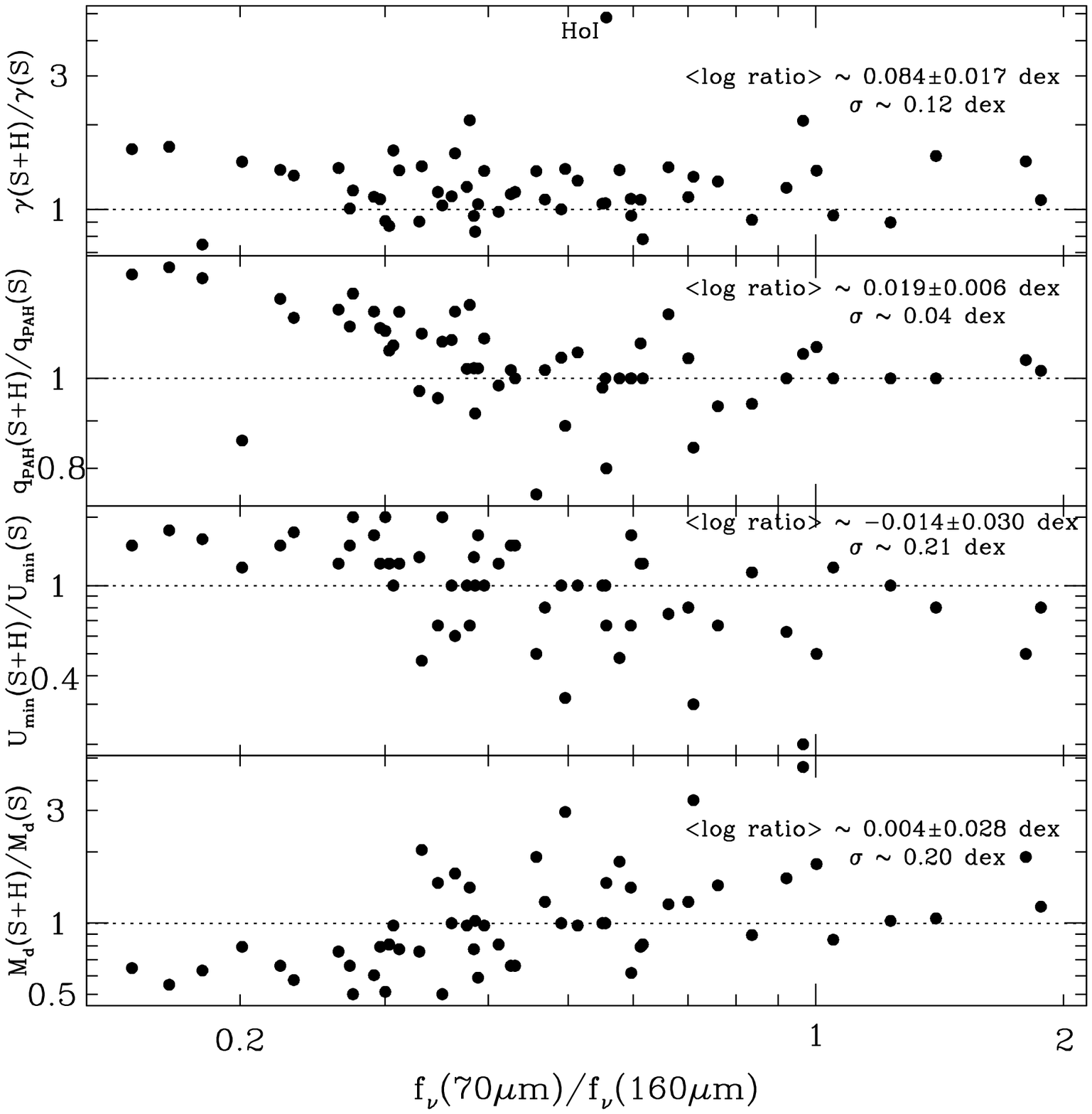} 
 \caption{Comparison of the dust model parameters obtained from fitting \cite{draineli07} spectral energy distribution models to the entire observed photometric data from 3.6 to 500\m, for fits including both Spitzer and Herschel data (`S+H') versus fits using just the Spitzer photometry (`S').  The parameters constrain the quantity of interstellar dust and their heating; see \S~\ref{sec:fits}.  The comparison is made as a function of the $f_\nu(70\mu m)/f_\nu(160\mu m)$ ratio, which is related to the average temperature of the interstellar dust grains.  Reference dotted lines are drawn for a ratio of unity.}
 \label{fig:ratios}
\end{figure}

\begin{figure} 
 \plotone{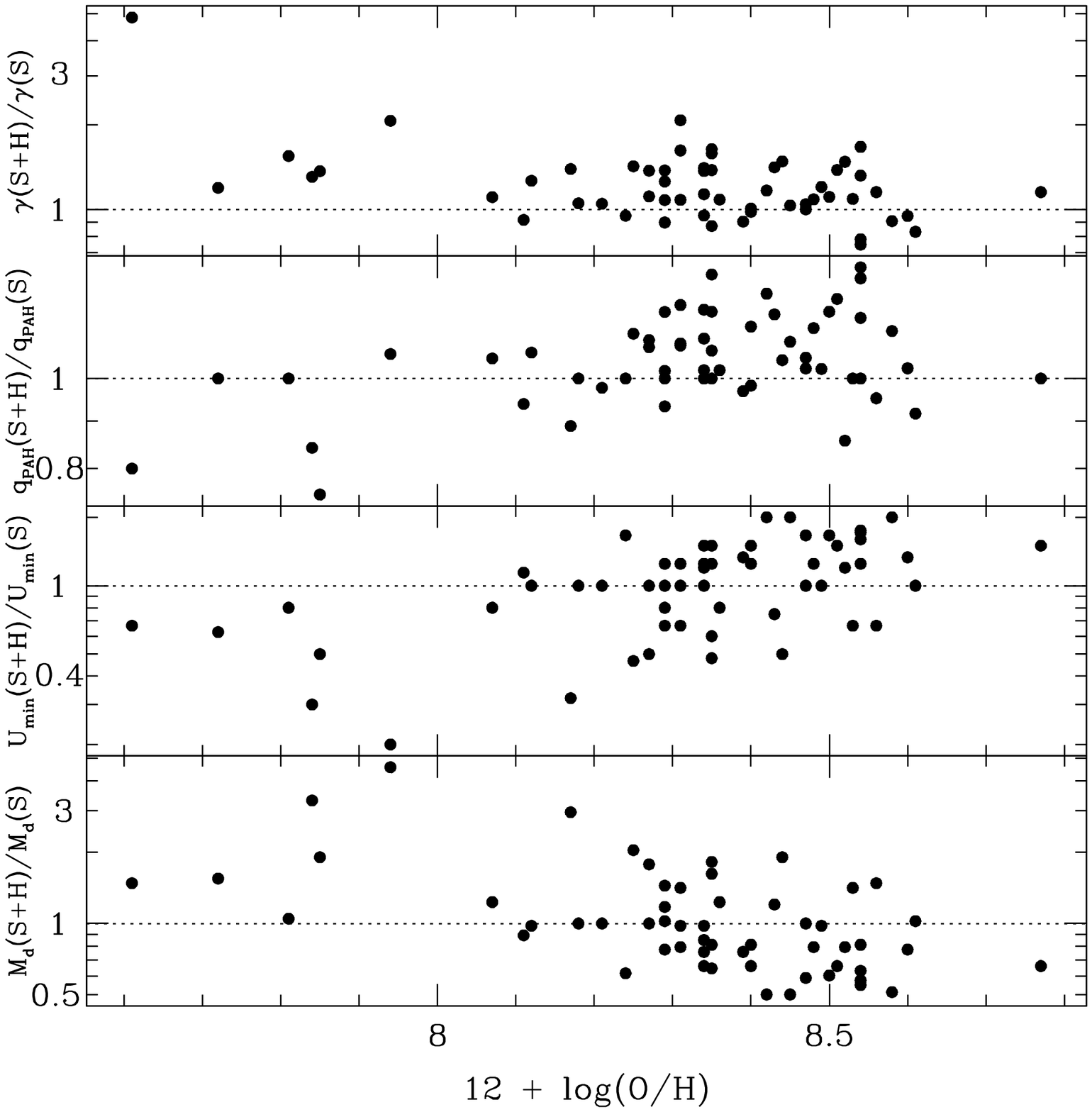} 
 \caption{The same as Figure~\ref{fig:ratios} except plotted as a function of oxygen abundance as presented in \cite{moustakas10}, using data placed on the \cite{pilyugin05} metallicity scale.}
 \label{fig:ratiosOxygen}
\end{figure}

\begin{figure} 
 \plotone{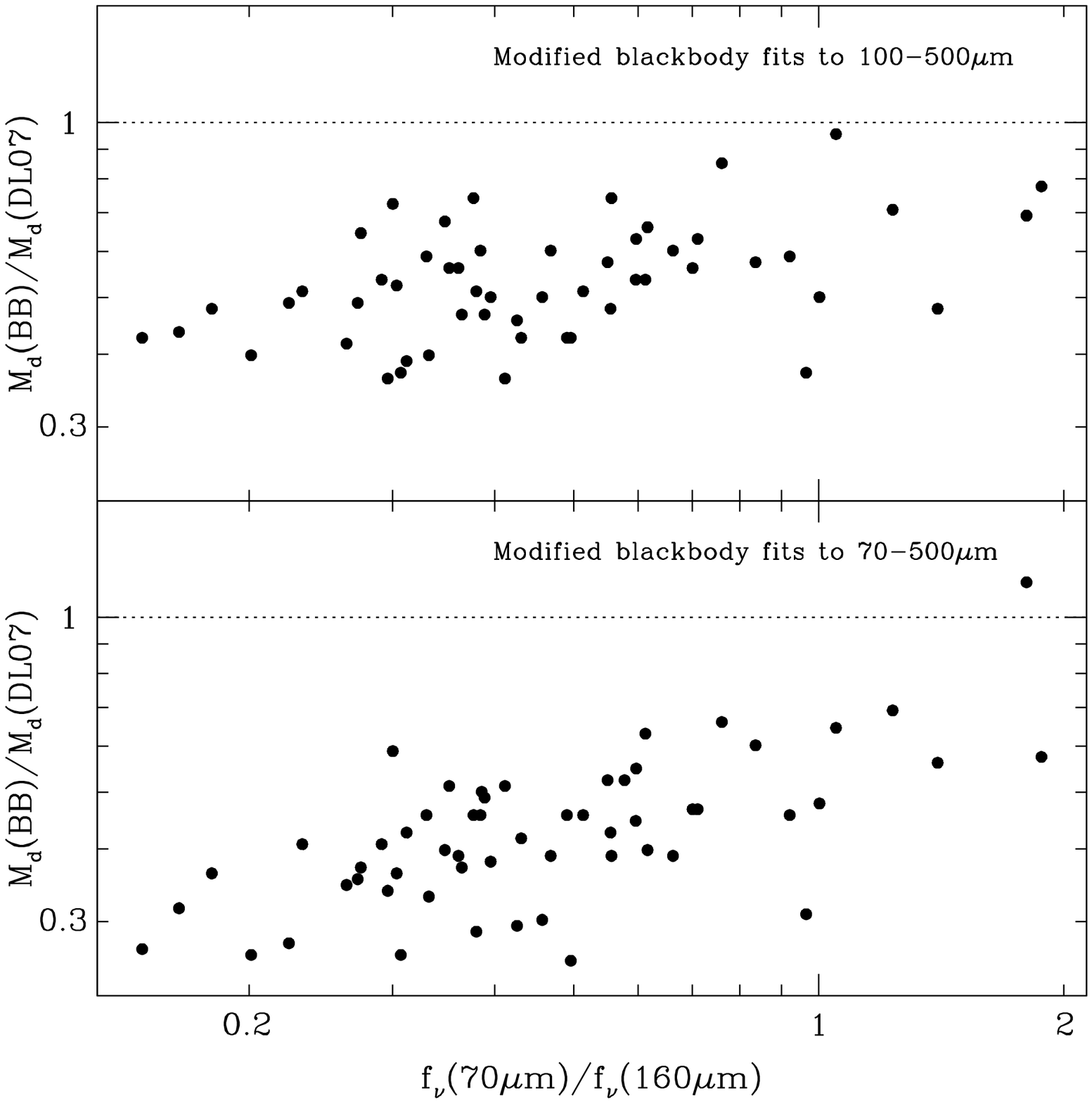} 
 \caption{The ratio of single-temperature, blackbody-based dust mass to that obtained from \cite{draineli07} model fits to the observed infrared/sub-millimeter spectral energy distributions.  Top (Bottom): The modified blackbody fits are based on the (error-weighted) Herschel 100, 160, 250, 350, and 500\m\ (70, 100, 160, 250, 350, and 500\m) photometry, and both $T_{\rm d}$ and the dust opacity coefficient $\beta$ are allowed to freely vary.  Reference dotted lines are drawn for a ratio of unity.}
 \label{fig:Mdust}
\end{figure}

\begin{figure} 
 \plotone{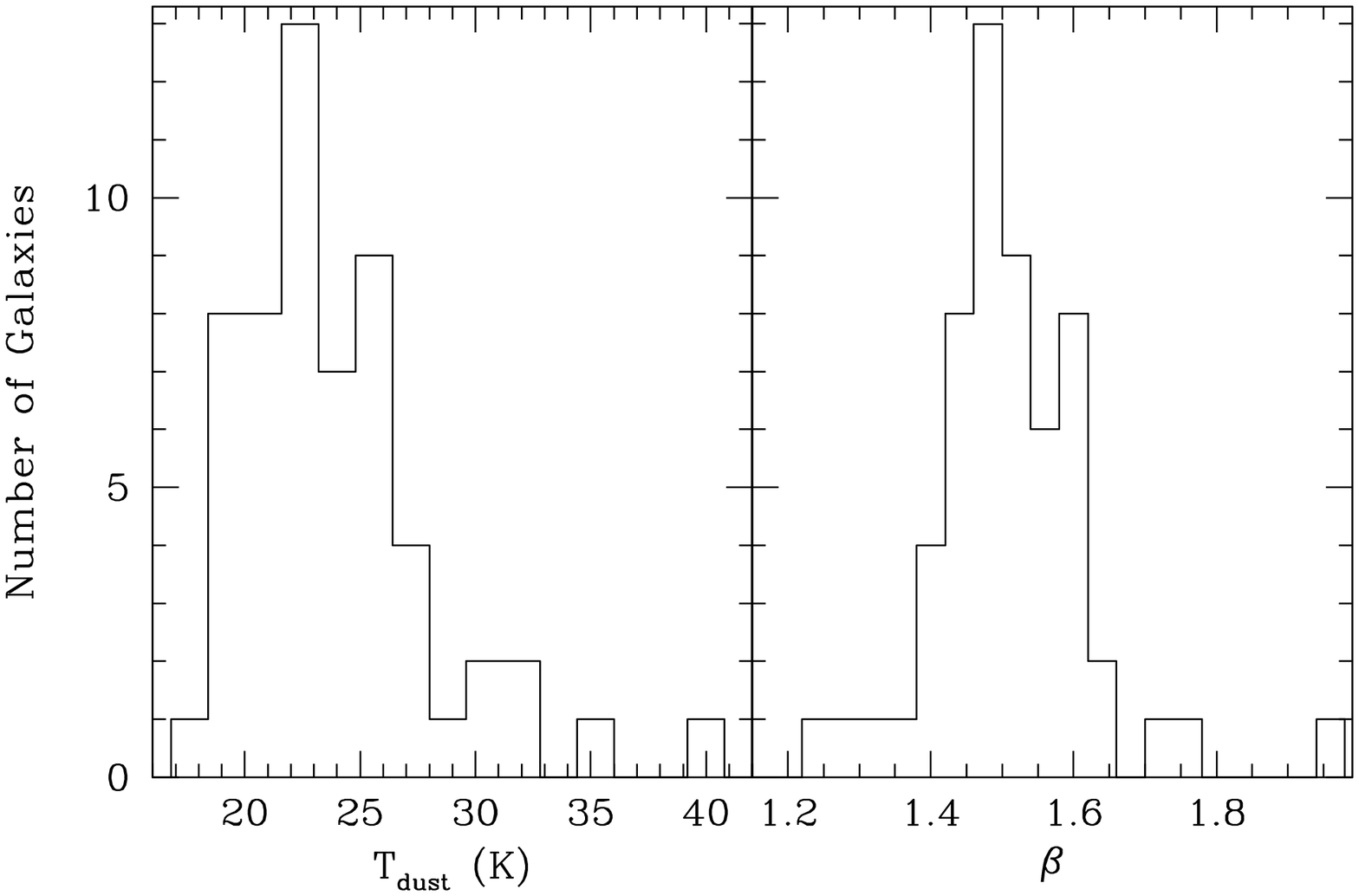} 
 \caption{The distributions of dust temperature and dust emissivity index $\beta$ when single-temperature modified blackbodies are fit to the (100--500\m) infrared spectral energy distributions.}
 \label{fig:TdustBeta}
\end{figure}

\begin{figure} 
 \plotone{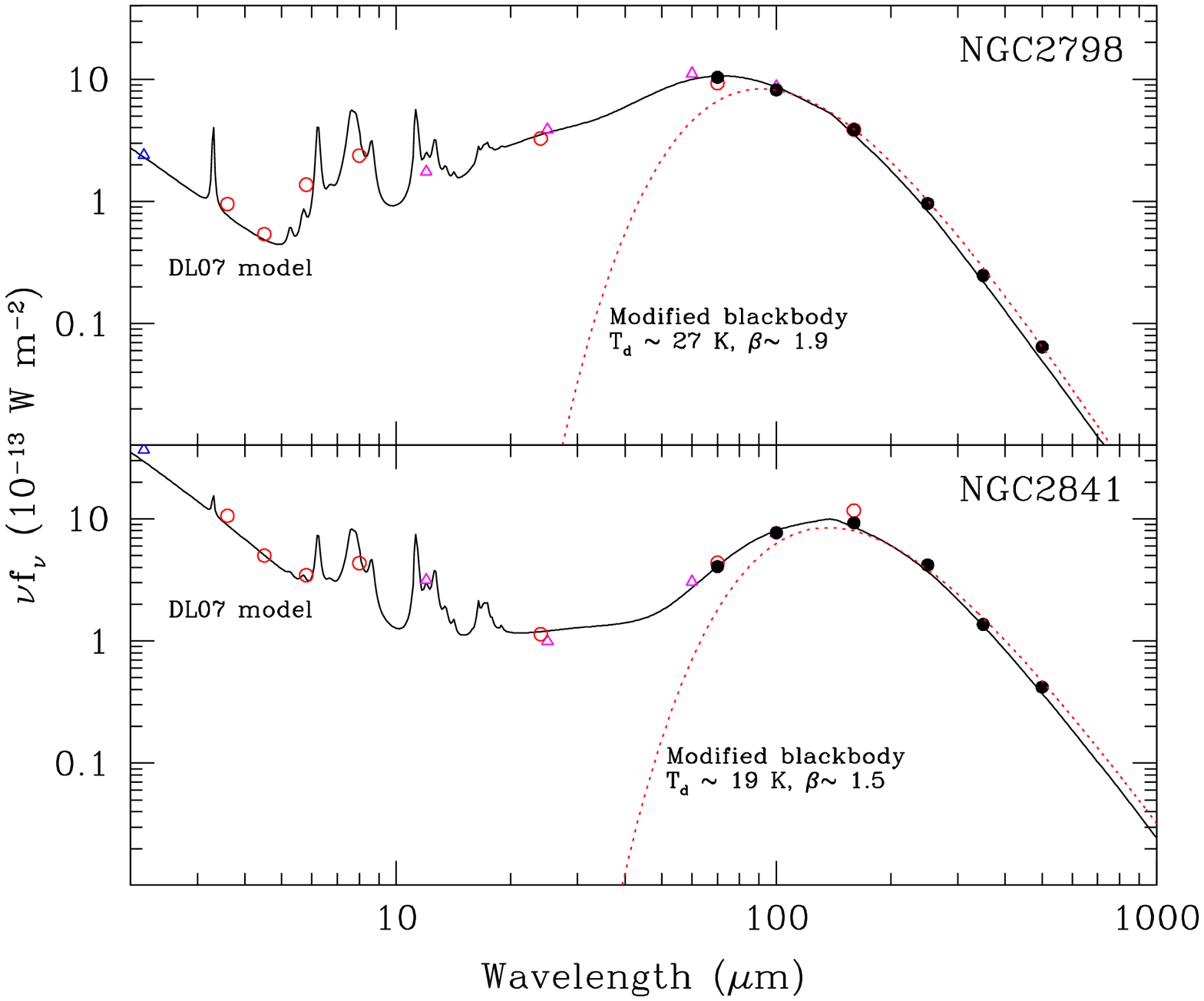} 
 \caption{A comparison of fitting \cite{draineli07} models to 3.6--500\m\ data versus fits of single temperature blackbodies to 100--500\m\ photometry, for a galaxy with warm dust (NGC~2798) and a galaxy with cool dust (NGC~2841).}
 \label{fig:Draine_v_BB}
\end{figure}

\end{document}